\documentclass[a4paper,11pt]{article}
\pdfoutput=1 

\usepackage{jcappub} 

\usepackage[T1]{fontenc} 

\usepackage{caption}
\usepackage[center]{subfigure}
\usepackage{multirow}
\usepackage{makecell}
\usepackage{booktabs}
\usepackage{float}
\usepackage{comment}
\usepackage{ulem}
\usepackage{booktabs}
\usepackage{enumerate}   
\usepackage{orcidlink}
\newcommand\scalemath[2]{\scalebox{#1}{\mbox{\ensuremath{\displaystyle #2}}}}

\title{\LARGE\boldmath 	
Probing regular black holes with sub-Planckian curvature through periodic orbits and their gravitational wave radiation
}

\author[a,b,1]{\large Soroush Zare,\orcidlink{0000-0003-0748-3386}\note{Corresponding author.}}
\author[c,d]{Tao Zhu,\orcidlink{0000-0003-2286-9009}}
\author[a]{Luis M. Nieto,\orcidlink{0000-0002-2849-2647}}
\author[c,d]{Shuo Lu,}
\author[a,e]{and Hassan Hassanabadi \orcidlink{0000-0001-7487-6898}}
 
\affiliation[a]{ Departamento de F\'{\i}sica Te\'orica, At\'omica y Optica and Laboratory for Disruptive \\ Interdisciplinary Science (LaDIS), Universidad de Valladolid, 47011 Valladolid, Spain}
\affiliation[b]{Helsinki Institute of Physics, University of Helsinki, P.O. Box 64, FI-00014 Helsinki, Finland}

\affiliation[c]{Institute for Theoretical Physics and Cosmology, Zhejiang University of Technology, Hangzhou, 310023, China}
\affiliation[d]{United Center for Gravitational Wave Physics (UCGWP), Zhejiang University of Technology, Hangzhou, 310023, China}

\affiliation[e]{Department   of   Physics,   University   of   Hradec   Kr\'{a}lov\'{e}, Rokitansk\'{e}ho   62,   500   03   Hradec   Kr\'{a}lov\'{e},   Czechia}

\emailAdd{szare@uva.es}
\emailAdd{zhut05@zjut.edu.cn}
\emailAdd{luismiguel.nieto.calzada@uva.es} 
\emailAdd{lushuo@zjut.edu.cn}
\emailAdd{hassan.hassanabadi@uva.es}

 \abstract{
 Extreme mass-ratio inspirals (EMRIs) are among the key targets for future space-based gravitational wave detectors. The gravitational waveforms emitted by EMRIs are highly sensitive to the orbital dynamics of the small compact object, which in turn are determined by the geometry of the underlying spacetime.
 In this paper, we explore the detectability of regular black holes with sub-Planckian curvature, which can be interpreted as regularized versions of the Schwarzschild black hole (RSBH). To do so, we begin by analyzing the metric and geodesics, determining the effective potential, and investigating the marginally bound orbits and the innermost stable circular orbits for timelike particles.
 Our analysis reveals that orbital radius, angular momentum, and energy significantly depend on the model parameter $\alpha$ for both orbits. 
 In addition, we study how variations in $\alpha$ influence the photon sphere and the corresponding shadow silhouette. Observations of M87* and Sgr A* motivate testing whether $\alpha$ falls within an observationally constrained range, narrower than the theoretical bound ensuring a singularity-free BH structure.
 Our main aim is to focus on the influence of the model parameter on a specific kind of orbit, the periodic orbit, surrounding a supermassive RSBH.  The findings show that, for a constant rational integer, $\alpha$ has a significant impact on the energy and angular momentum of the periodic orbit.
 Utilising the numerical kludge method, we further investigate the gravitational waveforms of the small celestial body over various periodic orbits.  The waveforms display discrete zoom and spin phases within a complete orbital period, influenced by the RSBH parameter $\alpha$. 
 As the system evolves, the phase shift in the gravitational waveforms grows progressively more pronounced, with cumulative deviations amplifying over time. 
 With the ongoing advancements in space-based gravitational wave detection systems, our results will aid in leveraging EMRIs to probe and characterize the RSBH properties.
}

\keywords{Schwarzschild-like black hole; astrophysical black holes;  periodic orbit; gravitational radiations}

\makeatletter
\gdef\@fpheader{}
\makeatother

\begin{document}
\maketitle
\flushbottom

\section{Introduction}
In recent years, black holes (BHs) and gravitational waves (GWs), two of the most remarkable predictions of general relativity (GR), have been confirmed to a high degree of confidence through a series of groundbreaking observational discoveries \cite{AbbottPRL2016-1,AbbottPRL2016-2,AbbottPRX2021,AkiyamaL12019,AkiyamaL52019,AkiyamaL62019,AkiyamaL122022,AkiyamaL172022}.
 Many high-precision experiments in gravitational physics, specifically the LIGO/Virgo gravitational wave (GW) detectors and long baseline interferometers such as the Event Horizon Telescope (EHT) \cite{AkiyamaL12019,AkiyamaL52019,AkiyamaL62019,AkiyamaL122022,AkiyamaL172022,PsaltisPRL2020} and the GRAVITY instrument of the European Southern Observatory \cite{GRAVITYAA2020,BaubockAA2020},  have significantly enhanced our comprehension of gravity in the strong-field regime.
These observational breakthroughs offer a powerful framework for probing numerous open questions in modern physics, as reviewed in Ref. \cite{BarackCQG2019}. 
Notably, these observations provide valuable insights into the longstanding hypothesis that complete gravitational collapse results in the formation of BHs \cite{OppenheimerPR1930}. 

The aforementioned observations appear to be in accordance with the hypothesis that the end state of complete gravitational collapse is a Kerr BH, fully characterized by its mass and angular momentum. Photons approaching a BH with a sufficiently small impact parameter will get trapped by the event horizon and will not reach an asymptotic observer, hence creating the phenomenon known as a shadow \cite{LuminetAA1979,FalckeApJL13,GrallaPRD2019,CunhaGRG2018,CunhaPLB2017,NarayanApJL2019,PerlickPR2022}, which has been empirically observed by the EHT collaboration.
Therefore, the study of BH shadows has attracted widespread attention, with many works focusing on how shadow observations can deepen our understanding of BH properties \cite{TsupkoPRD2017,BroderickApJ2022,MizunoNatAst2018} and shed light on possible deviations in the underlying spacetime geometry \cite{AtamurotovPRD2013,AbdujabbarovSS2016,AbdikamalovPRD2019,AtamurotovPRD2015,AtamurotovCPC2023,BelhajPLB2021,BelhajCQG2021,WeiJCAP2019,LingPRD2021,TsukamotoPRD2018,AraujoFilhoCQG2024,RayimbaevPoDU2022,PerlickPRD2018,RosaPRD2023,VagnozziCQG2023,KocherlakotaPRD2021}. Such deviations may arise either from additional parameters introduced in alternative theories of gravity \cite{UniyalPoDU2023,LambiaseEPJC2023,KhodadiPRD2022,KhodadiJCAP2021,PanotopoulosPRD2021,EslamPanahEPJC2024,Meng2025arxiv,Xu2024arxiv,Huang2025arxiv,FengAstpP2025,Waseem2025arxiv,Kala2025arxiv,Yin2025arxiv,ZareJCAP2024,StuchlikEPJC2019} or from the astrophysical environment surrounding the BH \cite{WuPoDU2024,PantigJCAP2022,ChenEPJP2024,XuJCAP2018,CapozzielloJCAP2023,CapozzielloPoDU2023,Nieto2025,SekhmaniJHEA2025,KonoplyaPLB2019,ZarePLB2024}.

Nonetheless, the collapse of matter into a BH inevitably results in the appearance of singularities, which are geodesically incomplete areas of spacetime indicating that the scalar curvature diverges near the BH's center \cite{PenrosePRL1965,Penrose1969,Hawking1966,Joshi2011,GoswamiPRL2006,JanisPRL1968}.
Notably, when the quantum effects of matter are considered, the BH may undergo evaporation via Hawking radiation, ultimately resulting in the information loss paradox \cite{HawkingPRD1976,Giddings1992,Hawking1975,Preskill,LiIJMPD2013,ChenPR2015,CasadioPLB2000}.
It is widely accepted that a complete resolution of these fundamental issues, and a deeper understanding of the associated puzzles, ultimately depends on a complete theory of quantum gravity \cite{DeWittPR1967,tHooftNPB1985,CallanPRD1992,DonoghuePRL1994,GarayIJMPA1995,HanPLB2005,CalmetEPJC77}. In the absence of such a theory, one possible approach to tackle the implications of the singularity theorems is to construct regular BH solutions that avoid singularities at a phenomenological level. This can be achieved, for example, by introducing some exotic matter that violates the classical energy conditions, or by incorporating quantum corrections to the spacetime geometry. Such modifications, that is, exotic matter or quantum geometry may arise from nonlinear electrodynamics within a classical framework, or from quantum gravitational effects that become relevant in the strong-curvature regime.
For comprehensive reviews, please see Refs. \cite{Ashtekar2023,LanIJTP2023}.  Traditionally, some of the most prominent examples of singularity-free regular BHs include the Bardeen, Hayward, and Frolov BHs \cite{Bardeen1968,HaywardPRL2006,FrolovJHEP2014}. These models are identified by a de Sitter (dS) core at the BH’s center, which replaces the traditional singularity.
A novel class of regular BHs was recently proposed by Ling and Wu \cite{LingCQG2023}, based on the  earlier work by Li, Ling, and Shen \cite{LiIJMPD2013}. In these models, the Newtonian potential is exponentially suppressed, leading to a spacetime geometry that approaches Minkowski space near the BH center, rather than exhibiting a de Sitter (dS) core. Notably, the Kretschmann scalar remains sub-Planckian throughout the entire spacetime.
Several other examples of regular BHs featuring a Minkowskian core can also be found in the literature; see, for instance, \cite{Culetu,CuletuIJTP2015,RodriguesPRD2016,SimpsonUniv2019,ZengPRD2023,TangEPJC2024,ZhangEPJC2024}
. Despite their differing core structures, both the Minkowskian-core and dS-core BHs exhibit similar asymptotic behavior at spatial infinity. A one-to-one correspondence between these two classes of solutions has been established in Ref. \cite{LingCQG2023}.



The promising detection of binary BH mergers has ushered in the exciting era of GW astronomy, marking a decisive step toward probing the strong-field dynamics of compact objects such as BHs through their GW signals. 
These detections, reported by the LIGO and Virgo collaborations \cite{AbbottPRL2016-1,AbbottPRL2016-2,AbbottPRX2021}, have not only verified the predictions of GR in extremely nonlinear regimes but have also opened new opportunities for exploring celestial bodies whose structure may be influenced by quantum gravity effects. 
Extreme mass ratio inspirals (EMRIs), featuring a stellar-mass compact object like a BH or neutron star steadily spiralling towards a supermassive BH, are some of the most intriguing targets for future space-based GW detectors. Space-based gravitational wave detectors such as LISA (i.e., Laser Interferometer Space Antenna) \cite{LISA1,LISA2}, Taiji \cite{HuNSR2017}, TianQin \cite{LuoCQG2016}, and DECIGO (i.e., DECi-heltz Gravitational-wave Observatory) \cite{DECIGO} are ideally designed to detect the prolonged, low-frequency signals produced by these systems. EMRIs hold particular importance as their gravitational waveforms encode intricate information regarding the spacetime configuration of the central BH \cite{HughesCQG2001,GlampedakisCQG2005-2,BarausseGRG2020}. This offers a unique opportunity to investigate the Kerr geometry anticipated by General Relativity and to explore possible deviations triggered by modern physics, including modified gravity theory or the existence of exotic compact objects \cite{CardosoLRR2019,BabakPRD2017}.

A key aspect of EMRI dynamics is the existence of periodic orbits--bounded trajectories wherein a test particle returns to its starting point after an integer number of radial and angular oscillations \cite{LevinPRD2008,LevinPRD2009}.  The orbits, labeled by zoom -- whirl parameters $(z; w; \nu)$ \cite{LevinPRD2008,LevinPRD2009,LevinCQG2009}, provide basic modes for comprehending generic orbital dynamics \cite{LevinPRD2008,LevinCQG2009,GrossmanPRD2009,MisraPRD2010,BabarPRD2017}.
Building on the taxonomy proposed in Ref. \cite{LevinPRD2008}, periodic orbits have been extensively investigated across a wide range of BH spacetimes. Examples include the charged BHs \cite{MisraPRD2010}, BH with a dark matter halo \cite{Alloqulov2025arxiv,HaroonPRD2025}, regular BHs \cite{GaoAP2020,MengEPJC2025}, BH free of Cauchy horizon \cite{WangJCAP2025}, naked singularities \cite{BabarPRD2017}, Kerr--Sen BHs \cite{LiuCTP2019}, polymer BHs in a modified gravity model \cite{TuPRD2023}, Einstein-Lovelock BHs \cite{LinPoDU2021}, quantum-corrected BHs \cite{DengPoDU2020,YangJCAP2025}, Kehagias-Sfetsos BHs \cite{WeiPRD2019}, and brane-world black holes \cite{DengEPJC2020}, and hairy BHs arising in Horndeski theory \cite{LinEPJC2023}. For further studies of periodic orbits in other BH models, we refer the reader to Refs. \cite{LinAP2023,ChanGRG2025,ZhouEPJC2020,ZhangEPJC2022,YaoPRD2023,QiEPJC2024,Li-KuangPRD2023,JiangPoDU2024,Lu2025arxiv,HuangPRD2025,ZhaoEPJC2024,ShabbirPoDU2025,LiEPJC2024, Wang:2025wob} and references therein.
Gravitational wave radiations from EMRIs encode distinctive imprints of periodic orbits, most prominently through the characteristic zoom–whirl phases that arise during the inspiral. This connection has inspired extensive investigations of Gravitational waveforms from periodic orbits in a variety of BH spacetimes and studies assessing their potential observability with future space-based detectors \cite{YangJCAP2025,WangJCAP2025,TuPRD2023,WeiPRD2019,HuangPRD2025,ZhaoEPJC2024,ShabbirPoDU2025,LiEPJC2024}.

It should be emphasized that the characteristics of geodesic motion are directly governed by the geometry of the BH spacetime. Recently, a new class of regular BHs has been proposed, characterized by sub-Planckian curvature, an asymptotically Minkowskian core, and an exponentially suppressed gravitational potential. The deviation from the standard Newtonian potential in these models can be interpreted as an effective manifestation of quantum-gravity–induced corrections to spacetime structure.
Relative to the singular BHs predicted by GR, the geometry of regular BHs exhibits notable differences. It is therefore reasonable to expect that periodic geodesic orbits will be modified by quantum-gravity corrections. Such quantum-corrected BHs \cite{LingCQG2023} have drawn considerable attention, with various aspects explored in the literature, including accretion disks \cite{ZengPRD2023}, quasinormal modes \cite{TangEPJC2024,ZhangEPJC2024}, and so on.

In this work, we examine the geodesic motion of particles around regular BHs with sub-Planckian curvature and a Minkowskian core, which may be regarded as a regularized Schwarzschild BH (RSBH). We begin by discussing key geometrical properties of the solution, including scalar invariants, before focusing on timelike periodic geodesic orbits around the proposed BH. The equations of motion for massive test particles are derived, and the effective potential is obtained from the corresponding Lagrangian \cite{Chandrasekhar}.
Particular attention is given to the role of the model parameter $\alpha$, which arises from quantum-gravity effects and encodes modifications to the standard Heisenberg uncertainty principle owing to quantum gravity. 
Periodic orbits and associated GW signals for regular BHs with sub-Planckian curvature and a Minkowskian core have not been systematically investigated in the literature.
Therefore, we explore how $\alpha$ influences the effective potential, the marginally bound orbits (MBOs), and the innermost stable circular orbits (ISCOs) \cite{Misner}. Furthermore, we investigate the dependence of the photon sphere and the associated shadow silhouette on variations of $\alpha$. Using observational data from M87* and Sgr A*, this analysis further motivates an investigation into whether $\alpha$ can be constrained to an observationally viable range, narrower than the theoretical interval required to ensure a singularity-free BH geometry.
In this work, we model the RSBH as the central supermassive BH and treat stellar-mass objects as test particles, thereby forming EMRIs. By studying the trajectories of periodic orbits, we investigate the associated GW radiations from EMRIs, offering insight into the radiation emitted in such systems. Our study highlights the sensitivity of the model parameter $\alpha$ to orbital radii and angular momenta, the allowed $L-E$ parameter space for bound orbits, and the $(z; w; \nu)$ taxonomy of periodic orbits around the central RSBH. Finally, employing the numerical kludge approach \cite{BabakPRD2007}, we simulate GWs from a stellar-mass object orbiting a supermassive regular BH with sub-Planckian curvature, and we discuss in detail the impact of the quantum-gravity parameter $\alpha$ on these signals.

The structure of this paper is as follows. In Sect. \ref{Sec2}, we briefly review the spacetime of regular black holes with sub-Planckian curvature and a Minkowskian core, and we then examine the associated scalar invariants. In Sect. \ref{Sec3}, we study the geodesic motion around such an RSBH solution by analyzing the effective potential and bound orbits for timelike particles, as well as the photon sphere and shadow silhouette for lightlike particles. Section \ref{POs} is devoted to the classification of zoom–whirl structures and bound orbits, characterized by the rational number q, which is defined through a triplet of integers describing the orbital structure around the central RSBH. Section \ref{GWs} addresses the gravitational-wave radiation associated with these dynamics. Finally, the results are summarized in Sect. \ref{SecC}.
Throughout this work, we adopt geometrized units, setting $G = c = 1$.

\section{Regularized Schwarzschild black hole}\label{Sec2}

In this section, we first give a brief review of the spherically symmetric regular BH that was put forward in \cite{LingCQG2023}, which features an asymptotically Minkowski core, an exponentially suppressing gravity potential, and sub-Planckian curvature. The corresponding metric is given by \cite{ZengPRD2023,TangEPJC2024,ZhangEPJC2024}
\begin{subequations}\label{metric}
	\begin{align}\nonumber
		ds^{2} & =g_{tt}dt^{2}+g_{rr}dr^{2}+g_{\theta\theta}d\theta^{2}+g_{\varphi\varphi}d\varphi^{2}\\ 
		&\equiv - A(r) dt^{2} + B(r)^{-1} dr^{2} + r^{2} (d\theta^{2}+\sin^{2}\theta d\varphi^{2}),\label{metric1}
	\end{align}
where the metric function $\text{A}(r)$ is of the form
	\begin{align}\label{MetricFuncA1}
	A(r) = B(r)=1+2\psi(r) = 1-\frac{2 M}{r} e^{-\alpha M^{x} r^{-p}}.
\end{align}
\end{subequations}
Where $\psi(r)$ denotes the modified Newton potential (i.e., an exponentially suppressing Newton potential), and $M$ is the BH mass. Here, $\alpha$, $p$, and $x$ are taken as dimensionless parameters \footnote{For convenience, we set $8\pi G = l_{p}^{2} = 1$ throughout this paper. 
Accordingly, if one wishes to restore the physical dimensions of the potential, it takes the form
$\psi = -\frac{MG}{r}\, e^{-\alpha M^{2/3} r^{-2}}$.
}. In particular, $\alpha$ denotes the degree of deviation from the Newtonian potential and defines the corrections that result from the effects of quantum gravity \cite{XiangIJMPD2013,LiAP2018}. 
Clearly, when $\alpha = 0$, the potential simplifies to the standard Newtonian potential, and this regular BH reverts to a traditional Schwarzschild BH. Moreover, in order to ensure that the spacetime curvature remains sub-Planckian,  the parameters should fulfil the requirements $p \geqslant x \geqslant p/3$ and $p \geqslant 2$ \cite{LingCQG2023}.
Without sacrificing generality, we focus on a regular BH metric with $x = 2/3$ and $p = 2$, which exhibits the same asymptotic behavior at spatial infinity as the Bardeen BH.
However, it is well-known that these two BH models exhibit different behaviors near the central region. In the limit $r \to 0$, the regular BH with sub-Planckian curvature (for $x = 2/3$ and $p = 2$) approaches a Minkowskian core, whereas the Bardeen BH develops a de Sitter core.
Ensuring that Eq. \eqref{metric} describes a BH -- namely, one with a horizon and free of singularities -- the deviation parameter must lie within the interval $\alpha \in [0, 2 M^{4/3}/e]$, with $e$ being the Euler’s number. For $0 \leq \alpha < 2 M^{4/3}/e$, the RSBH has both Cauchy and event horizons (corresponding to a non-extreme BH), which merge into a single degenerate horizon (corresponding to an extreme BH) when $\alpha = 2 M^{4/3}/e$. For $\alpha > 2 M^{4/3}/e$, no horizon exists.
In fact, the parameter $\alpha$ describes the regularization of the Schwarzschild BH. The limit as $\alpha$ approaches $0$ eliminates regularization and recovers the Schwarzschild BH solution. For a large value of $r$, we evidently recover the Schwarzschild BH geometry. Thus, one can see that the regularized Schwarzschild BH (RSBH) spacetime is asymptotically flat. 
At a large scale where $r\gg \sqrt{\alpha}M^{1/3}$, the metric function $A(r)$ can be expressed in the following expansion 
\begin{equation}\label{MetricFuncA2}
	A(r) = 1-\frac{2M}{r}\left(1-\frac{\alpha M^{2/3}}{r^{2}}+\frac{1}{2}\left(\frac{\alpha M^{2/3}}{r^{2}}\right)^{2}-\frac{1}{6}\left(\frac{\alpha M^{2/3}}{r^{2}}\right)^{3}+
	\mathcal{O}\left(-\frac{\alpha M^{2/3}}{r^{2}}\right)^{4}\right).
\end{equation}

\begin{figure}[htb]
\centering 
\includegraphics[width=.49\textwidth]{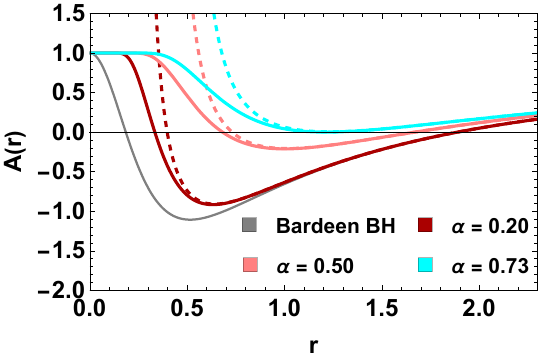}
	\caption{\label{sec2-fig1}
    Plot of the metric function $A(r)$ for different values of $\alpha$. The comparison includes the Bardeen BH, the full RSBH solution (solid curves), and its asymptotic approximation (dashed curves). }
\end{figure}

Indeed, all regular BHs may be expressed as analogous polynomials, indicating that the regular metric asymptotically resembles that of the Schwarzschild BH. Figure \ref{sec2-fig1} shows the distinction between the metrics of the Bardeen BH, RSBH and the asymptotic approximation of the RSBH spacetime. The RSBHs exhibiting lower $\alpha$, which is the deviation parameters in RSBH spacetime, would closely resemble Bardeen BH spacetime, particularly around the BH horizon, except in the vicinity of the single point $r = 0$.
As $r\rightarrow\infty$, the metric in Eq. \eqref{metric} and in Eq. \eqref{MetricFuncA2} reduces to the Schwarzschild BH. For simplicity, in all plots, we take $M=1$. In Fig. \ref{sec2-fig1}, it can be seen that with changing $\alpha$, the function $A(r)$ will change.

\begin{figure}[htb]
	\centering 
	\includegraphics[width=.49\textwidth]{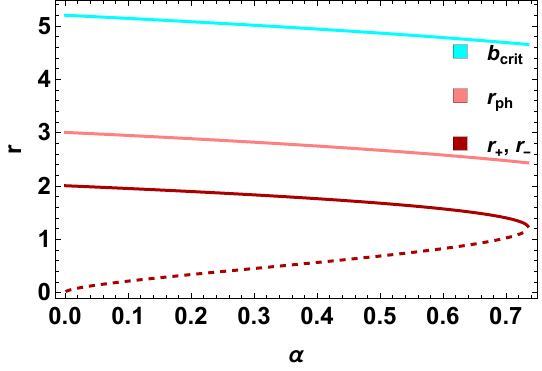}
	\caption{\label{sec2-fig2} 
    The Plot indicates the dependence of the different radii and the critical impact parameter on the variable $\alpha$. The event horizon radius (Red solid line), Cauchy horizon radius (Red dashed line), photon sphere radius (Pink solid line), and the critical impact parameter (Cyan solid line) are all plotted.}
\end{figure}

The metric \eqref{metric1} with lapse function $A(r)$ given in Eq. \eqref{MetricFuncA1} exhibits a coordinate singularity at 
\begin{equation}\label{Horizons}
A(r)=0,
\end{equation}
allowing for up to two real positive roots, denoted as $r_{\pm}$. 
Here, $r_{+}$ stands for the event (outer) horizon of the BH, whereas $r_{-}$ denotes the Cauchy (inner) horizon. By solving Eq. \eqref{Horizons} numerically, the locations of these horizons can be determined for different values of the parameter $\alpha$. This, as mentioned above, reveals a critical threshold for the existence of the event horizon at approximately $\alpha \lesssim 0.735$. 
Using numerical plotting on Eq. \eqref{Horizons},
Fig.~\ref{sec2-fig2} shows how both the event and Cauchy horizons vary within the RSBH geometry as functions of the parameter space $\alpha$. 
Except in the vicinity of the singular, that is, $r\rightarrow 0$, the RSBH solution \eqref{MetricFuncA1}, that is,the regular BH solution with sub-Planckian curvature and Minkowskian core clearly resembles its approximate asymptotic form \eqref{MetricFuncA2}, especially when the value of the model parameter $\alpha$ is smaller. However, the RSBH solution \eqref{MetricFuncA1} converges to its approximated form as $r \rightarrow \infty$.

\begin{figure}[htb]
	\centering 
	\includegraphics[width=.49\textwidth]{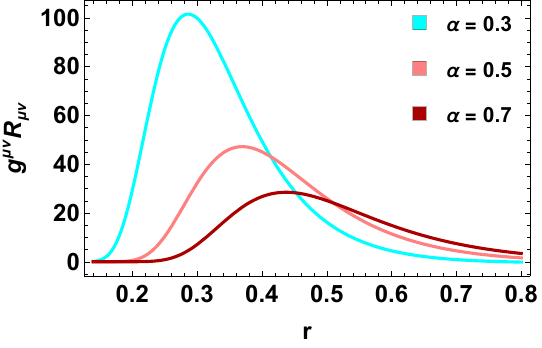}  
	\hfill
	\includegraphics[width=.49\textwidth]{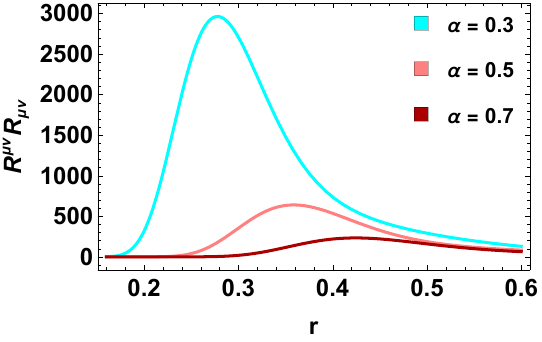}
	\hfill
	\includegraphics[width=.49\textwidth]{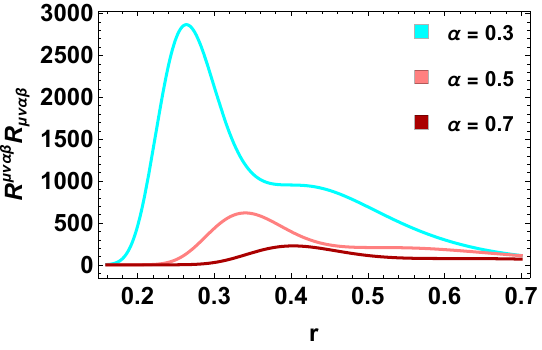}
	\caption{\label{fig1-3} 
Behaviour of scalar invariants in RSBH spacetime as a function of $r$ for varying values of the regularization parameter $\alpha$. Panels from left to right correspond to the Ricci scalar, to the Squared Ricci tensor, and to the Kretschmann scalar, respectively.
}
\end{figure}

The curvature invariants are quantities that elucidate the properties of the spacetime of a geometrical entity such as a BH. Notable scalar invariants encompass the Ricci scalar, the square of the Ricci tensor, and the Kretschmann scalar, which is the square of the Riemann curvature tensor. Here, we determine and discuss these quantities through investigating their graphical behaviors using Fig.~\ref{fig1-3}.
As a preliminary remark, let us note a useful mathematical detail: for any polynomial function $F(r)$, the ratio
$e^{-\alpha r^{-2}}/F(r) \;\to\; 0 \text{ as } |r| \to 0$. In other words, the exponential term dictates the asymptotic behavior in the small-$r$ limit.

Given the lapse function \eqref{MetricFuncA1}, we now proceed to examine the corresponding curvature invariants.

Let us first consider the Ricci scalar for the RSBH, which can be found as follows
\begin{equation}\label{RicciScalar}
R=g^{\mu\nu}R_{\mu\nu}= \frac{4 \alpha M^{5/3} \left(2 \alpha M^{2/3} - r^{2}\right)}{r^{7}}\, e^{-\alpha M^{2/3} r^{-2}}.
\end{equation}
It is obvious that when $\alpha = 0$, the Ricci scalar likewise equals zero. Hence, we explore the impact of the regularization parameter $\alpha$ on this scalar invariant numerically.  In the left panel of Fig. \ref{fig1-3}, we show \eqref{RicciScalar}, indicating that 
for regular BHs with sub-Planckian curvature and a Minkowskian core, the Ricci scalar $R$ remains finite and smooth in the limit $r \to 0$, hence it is well-defined in proximity to the central point $r = 0$, so confirming that the BH solution is regular in the center $(r \rightarrow 0)$. 

The square of the Ricci tensor $\mathcal{R}$ is the next quantity to be examined. It is provided for the spacetime metric in Eq. \eqref{metric} as
\begin{equation}\label{squareRicciTensor}
\scalemath{0.95}{
\mathcal{R}=R^{\mu\nu}R_{\mu\nu}= \frac{8 \alpha^{2} M^{10/3} \left(13 r^{4} - 12\alpha M^{2/3} r^{2} +4 \alpha^{2} M^{4/3}\right)}{r^{14}}\, e^{-2 \alpha M^{2/3} r^{-2}}
}.
\end{equation}
In the case of $\alpha = 0$, the Squared Ricci tensor is zero. We show the numerical plot analysis of Eq. \eqref{squareRicciTensor}. The middle panel of Fig. \ref{fig1-3} shows the quantity and its well-defined trend at the central point $r=0$. It should be noted that the Ricci scalar and the square of the Ricci tensor have the same characteristics.

Another scalar invariant that gives further information about the spacetime's curvature is the Kretschmann scalar, which, in particular, does not vanish in a Ricci flat spacetime \cite{RayimbaevPoDU2022}. 
For our case, it remains finite everywhere and vanishes in the limit $r\to 0$. By the way the Kretschmann scalar \cite{CapozzielloJCAP2023,CapozzielloPoDU2023} for 
regular BH with sub-Planckian curvature and a Minkowskian core becomes
 \cite{CapozzielloJCAP2023,CapozzielloPoDU2023}
\begin{equation}\label{KretschmannScalar}
\begin{split}	
K &=R^{\mu\nu\alpha\beta}R_{\mu\nu\alpha\beta} \\
&= \frac{16 M^{2} \left(3 r^{8} - 14\alpha M^{2/3} r^{6}+33\alpha^{2}M^{4/3}r^{4} -20 \alpha^{3} M^{2}r^{2}+4M^{8/3}\alpha^{4}\right)}{r^{14}}\, e^{-2 \alpha M^{2/3} r^{-2}}.
\end{split}
\end{equation}
In Fig. \ref{fig1-3}, we see that when $\alpha = 0$, the Kretschmann scalar is equal to $\frac{48M^{2}}{r^{6}}$. This matches the Kretschmann scalar for the Schwarzschild BH.
The features of the Kretchmann scalar are analogous to those of other scalar invariants previously addressed. 

Thus, as $r \to +\infty$, we find $e^{-\alpha r^{-2}} \to 1$, and all curvature invariants reduce to Laurent polynomial functions of $r$, asymptotically behaving as $\mathcal{O}(r^{-n})$. Consequently, the invariants vanish at large $r$, consistent with the spacetime approaching Minkowski space at infinity. In the opposite limit, as $r \to 0^{+}$, the invariants again approach zero, scaling as $\mathcal{O}(r)$, which indicates that the BH core is also asymptotically Minkowskian. Thus, the geometry approaches flat spacetime both at the center and at spatial infinity, with a finite region in between where the curvature attains its maximum \cite{SimpsonUniv2019}. We therefore conclude that all curvature invariants remain globally finite, confirming the absence of curvature singularities. In this sense, the geometry indeed represents a regular black hole spacetime, in the spirit of the construction originally proposed by Bardeen.

\section{Geodesics around regularized Schwarzschild black hole}\label{Sec3}

In this section, we examine the geodesic motion around the regular BH with sub-Planckian curvature and a Minkowskian core by analyzing the effective potential and bound orbits for timelike particles, and by investigating the photon sphere and shadow silhouette for lightlike particles. We are further motivated to investigate whether $\alpha$ falls within an observationally constrained range, narrower than the theoretical interval $\alpha \in [0,\, 2 M^{4/3}/e]$. Such a restriction would ensure that Eq. \eqref{metric} describes an RSBH, thereby establishing it as a viable candidate for astrophysical BHs. Throughout, particular attention is given to the role of the model parameter $\alpha$, which directly influences the geodesic structure of this static, spherically symmetric BH solution.
\subsection{Time-like geodesic: effective potential, ISCO and MBO radii}
The Lagrangian of a test particle’s geodesic motion governed by the RSBH spacetime \eqref{metric}  can be written as
\begin{equation}\label{Lagrangian}
2\mathcal{L}(x,\dot{x})=g_{\mu\nu}\dot{x}^{\mu}\dot{x}^{\nu}=\delta,
\end{equation}
where $g_{\mu\nu}$ refers to the spacetime metric, and a dot in the equation signifies differentiation with respect to the affine parameter. Here, $\delta$ for timelike particles is set to $-1$, whereas for lightlike particles it is set to $0$. 
Due to the isotropic gravitational field, the massive particle motion can be restricted to the equatorial plane, $\theta =\pi/2$, without any loss of generality. Based on the Lagrangian, the generalized momentum $p_{\mu}$ of the particle is given by $p_{\mu}=\partial \mathcal{L}/\partial \dot{x^{\mu}}$. Thus, the generalized momentum components can be found as follows 
\begin{subequations}\label{gmomentum}
\begin{align}
&p_{t}=g_{tt}\dot{t}=-E,\\
&p_{r}=g_{rr}\dot{r},\\
&p_{\varphi}=g_{\varphi\varphi}\dot{\varphi}=L.
\end{align}
\end{subequations}
Given that the metric is independent of $t$ and $\varphi$, their conjugate momenta $(p_{t},p_{\varphi})$ yield two associated constants, $E$ and $L$, which define the conserved energy and the conserved orbital angular momentum per unit mass, respectively. By combining the aforementioned three equations with the normalization requirement for the four-velocity of a massive particle, $\delta=-1$, we get:
\begin{subequations}
\begin{align}
&\dot{t}=-\frac{E}{A(r)}, \label{tdot}\\
&\dot{\phi}=\frac{L}{r^2}, \label{phidot}\\
&\dot{r}^2=E^2-A(r)\left(1+\frac{L^2}{r^2}\right) \label{rdot1}.
\end{align}
\end{subequations}

\subsubsection{Effective potential and bounded orbtis}
Motivated by studying periodic orbits around an RSBH, we will focus on these orbits, which represent a subset of bound orbits, constrained within a certain range of energy and angular momentum. Such bound orbits may be either stable or unstable.
An essential tool for figuring out orbital stability is the effective potential. Unstable orbits correspond to maxima, and stable orbits appear at minima of potential. Only the particle motion in the radial $r$ direction and angular $\varphi$ direction needs to be considered because the particle is on the equatorial plane. To analyze the effective potential, we can express radial Eq. \eqref{rdot1} as follows:
\begin{equation}\label{rdot2}
\dot{r}^2=E^2-V_{\mathrm{eff}},
\end{equation}
where $V_{\mathrm{eff}}$ represents the effective potential
\begin{equation}\label{effpot}
V_{\mathrm{eff}}=A(r)\left(1+\frac{L^2}{r^2}\right).
\end{equation}

The effective potential $V_{\text{eff}}(r)$, which is a function of the radial coordinate $r$ and is parameterized by $\alpha$, plays a key role in understanding the dynamics around the RSBH. Analyzing bound orbits of timelike particles is important for gaining insight into spacetime geometry. 
The properties of particle motion around the BH can be qualitatively understood by examining the profile of the effective potential in the spirit of Refs. \cite{Straumann,LiEPJC2024}. When $0 < L < L_{\text{crit}}$, where $L_{\text{crit}}$ denotes the critical angular momentum at which $V_{\text{eff}}$ develops a single extremum corresponding to the innermost stable circular orbit (ISCO, to be discussed later), the effective potential decreases monotonically from unity at infinity to zero at the horizon without developing a minimum. In this regime, no bound orbits exist. For $L > L_{\text{crit}}$, however, $V_{\text{eff}}$ acquires both a minimum, $V_{\min}$, and a maximum, $V_{\max}$, thereby allowing the possibility of bound orbits.
If $V_{\max} > 1$, particles with energies in the range $1 < E < V_{\max}$ may escape to infinity, corresponding to scattering orbits, while those with $V_{\min} < E < 1$ remain bounded. In contrast, when $V_{\max} < 1$, scattering orbits are excluded, and particles with $V_{\min} < E < V_{\max} < 1$ can form bound orbits. Since our interest lies in bound motion rather than scattering, we will restrict attention to the case $V_{\max} \leq 1$, with the special case $V_{\max} = 1$ corresponding to the marginally bound orbit (MBO). A more detailed analysis of these cases will be presented in the following.

According to Eq. \eqref{rdot2}, the particle's bound orbits are determined by two turning points where $\dot{r}^2 = 0$ is satisfied. One of these corresponds to the "Marginally Bound Orbits (MBO)," which occur when $E = 1$. The other, with $E < 1$, is known as the "Innermost Stable Circular Orbit (ISCO). 
For particles with $E > 1$, i.e., with energies exceeding that of the MBO, the motion is unbound and the particle can escape entirely from the gravitational field of the BH.
Therefore, for a test massive particle around a spherically symmetric BH spacetime (meaning that the spacetime geometry is the same in all directions, regardless of the angle), bounded orbits occur between MBOs and ISCOs. 

Between these two points, there exists a range of energies that allow for bound orbits to exist. 
The orbital angular momentum and energy of a particle in a bound orbit around a BH needs to satisfy 
\begin{equation}\label{LEcondition}
L_{\rm ISCO}\leq L \qquad \text{and} \qquad E_{\rm ISCO} \leq E \leq E_{\rm MBO} = 1,
\end{equation}
where $L_{\rm ISCO}$ being the orbital angular momentum of the particle along the ISCO, $E_{\rm MBO}=1$ corresponds to the energy of the MBO, where the particle has just enough energy to escape to infinity. $E_{\rm ISCO}$ represents the energy of the ISCO, and any particle with energy less than this will plunge into the black hole. The condition for angular momentum is slightly different. The angular momentum $L$ of a particle in a bound orbit must satisfy $L \ge L_{\rm ISCO}$, where $L_{\rm ISCO}$ is the angular momentum of the particle in ISCO.

\subsubsection{Marginally bound orbits}

\begin{figure}[htb]
\centering 
\includegraphics[width=.495\textwidth]{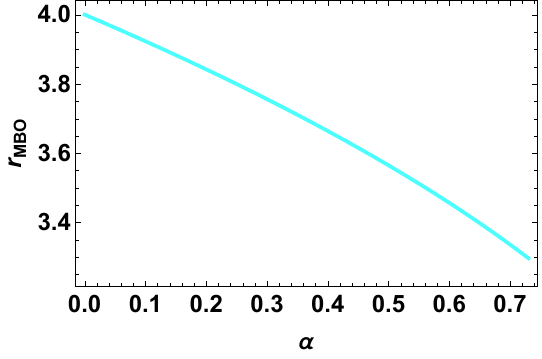} 
\hfill
\includegraphics[width=.495\textwidth]{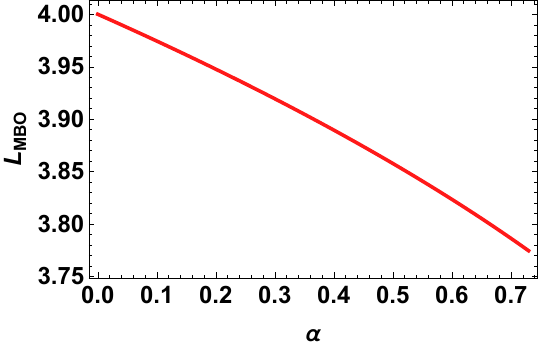} 
\caption{\label{fig3-4} 
The radius $r_{\text{MBO}}$ (left panel) and angular momentum $L_{\text{MBO}}$ (right panel) of MBOs around the RSBH as functions of the deviation parameter $\alpha$.}
\end{figure}

In the case of the MBO, a test particle possesses the same energy as a particle at rest at spatial infinity. This implies that no energy is lost or gained as the particle moves from infinity to the MBO, highlighting a state of energy conservation. The MBO corresponds to the smallest circular bound orbit with the minimum possible radius and a total energy $E_{\text{MBO}} = 1$. Based on the effective potential $V_{\text{eff}}$ given in Eq. \eqref{effpot}, this kind of the bound orbits, that is, the MBO must satisfy the conditions
\begin{equation}\label{MBOconditions}
V_{\mathrm{eff}}=E^2=1,\qquad\partial_{r}V_{\mathrm{eff}}=0.
\end{equation}
Using the above conditions, we can numerically examine the radius and orbital angular momentum of the MBO within the parameter range $\alpha \in [0, 0.735]$, while keeping the model parameter $p$ fixed at $p=2$. From Fig. \ref{fig3-4}, it can be observed that the particle’s radial distance $r_{\rm MBO}$ and its angular momentum $L_{\rm MBO}$ decrease with the BH parameter $\alpha$ in the RSBH model.

\subsubsection{Innermost stable circular orbits}
The ISCO represents the closest stable orbit for a timelike particle around a BH without plunging into the event horizon. Any disturbance inside this orbit would cause an inner spiral into the BH. The ISCO can be identified by the following requirements

\begin{equation}\label{IscoCond}
V_{\rm{eff}}=E^2,\qquad\partial_{r}V_{\rm{eff}}=0,\qquad\partial_{r,r}V_{\rm{eff}}=0.
\end{equation}
Let us now find the radius $r_{\rm{ISCO}}$ of ISCO for this scenario and see how the model parameter $\alpha$ affects the orbital movements by solving the three conditions mentioned above using the effective potential \eqref{effpot}, which includes $\alpha$.
One can derive the energy and angular momentum of the particle by using the first two conditions in Eq. \eqref{IscoCond}. However, the third condition allows us to precisely locate the ISCO, denoted as $r_{\text{ISCO}}$. Then, by substituting $r_{\text{ISCO}}$ into the previously derived equations, one can determine the corresponding energy $E_{\text{ISCO}}$ and angular momentum $L_{\text{ISCO}}$ of the orbit. Thus, we have
\begin{gather}\label{rLE-Isco}
r_{\rm{ISCO}}=\frac{3 A\left(r_{\rm{ISCO}}\right) A^{\prime}\left(r_{\rm{ISCO}}\right)}{2 A^{\prime 2}\left(r_{\rm{ISCO}}\right)-A\left(r_{\rm{ISCO}}\right) A^{\prime \prime}\left(r_{\text {ISCO}}\right)}, \\
L_{\rm{ISCO}}=r_{\rm{ISCO}}^{3 / 2} \sqrt{\frac{A^{\prime}\left(r_{\rm{ISCO}}\right)}{2 A\left(r_{\rm{ISCO}}\right)-r_{\rm{ISCO}} A^{\prime}\left(r_{\rm{ISCO}}\right)}},\\
E_{\rm{ISCO}}=\frac{A\left(r_{\rm{ISCO}}\right)}{\sqrt{A\left(r_{\rm{ISCO}}\right)-r_{\rm{ISCO}} A^{\prime}\left(r_{\rm{ISCO}}\right)/2}}.
\end{gather}

\begin{figure}[htb]
	\centering 
	\includegraphics[width=.495\textwidth]{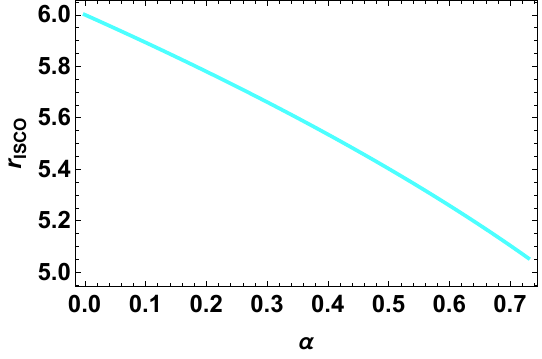} 
	\hfill
	\includegraphics[width=.495\textwidth]{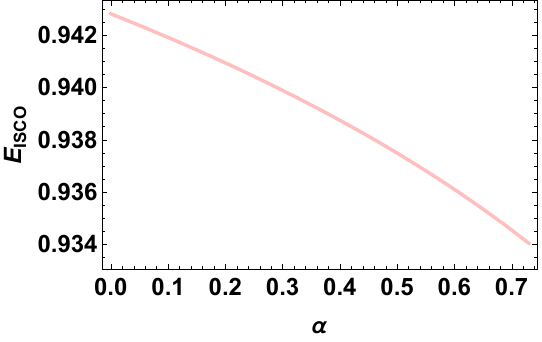} 
	\hfill
	\includegraphics[width=.495\textwidth]{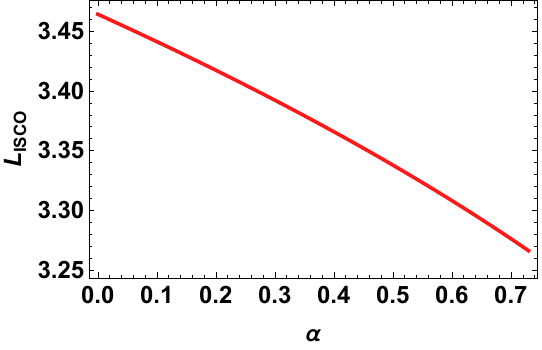} 
	\caption{\label{fig3-5} 
		The radius $r_{\text{ISCO}}$ (first panel), energy $E_{\text{ISCO}}$ (second panel), and angular momentum $L_{\text{ISCO}}$ (third panel) of ISCOs around the RSBH as functions of the deviation parameter $\alpha$.}
\end{figure}
Figure \ref{fig3-5} shows the numerical plots for $r_{\rm ISCO}$, $E_{\rm ISCO}$, and $L_{\rm ISCO}$ as functions of the RSBH model parameter $\alpha$. Their behavior is qualitatively similar to that in Fig. \ref{fig3-4}, deviating increasingly from the Schwarzschild case as $\alpha$ grows. For comparison, the Schwarzschild values are $r_{\rm ISCO} = 6M$, $E_{\rm ISCO} = 0.9428M$, and $L_{\rm ISCO} = 3.4641M$.

We next analyze the fundamental characteristics of bound orbits located between the MBO and ISCO around the RSBH, primarily based on the effective potential and the radial motion of a timelike particle. From the preceding discussion, it follows that for a fixed value of $\alpha$, the angular momentum of bound orbits is constrained to lie between the corresponding $L_{\rm ISCO}$ and $L_{\rm MBO}$.

\begin{figure}[htb]
	\centering 
	
	\includegraphics[width=.495\textwidth]{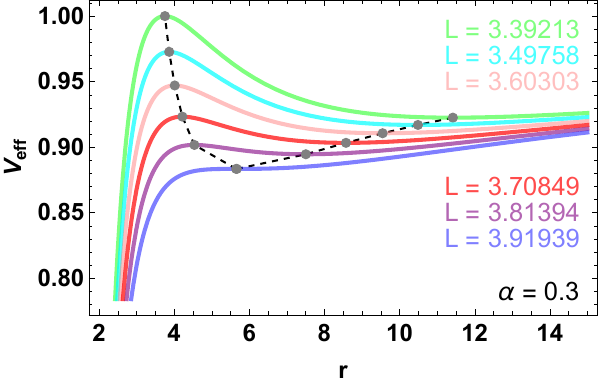} 
	\hfill
	\includegraphics[width=.495\textwidth]{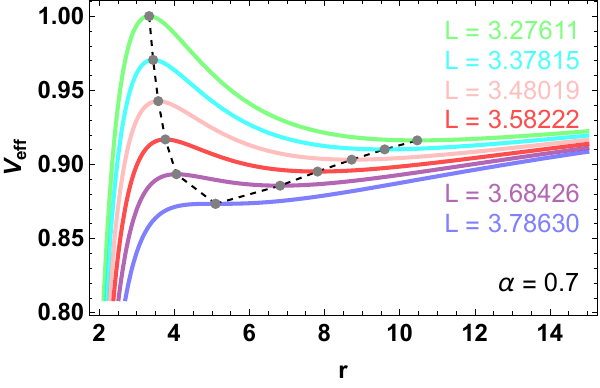} 
	\caption{\label{veff} 
		Effective potential $V_{\rm eff}$ as a function of $r$ for $\alpha = 0.3$ (left panel) and $\alpha = 0.7$ (right panel). The sequence of curves, from bottom to top, corresponds to angular momentum values increasing from $L_{\text{ISCO}}$ to $L_{\text{MBO}}$. The dashed lines mark the extrema of the effective potential.
	}
\end{figure}

\begin{figure}[htb]
	\centering 
	\includegraphics[width=.495\textwidth]{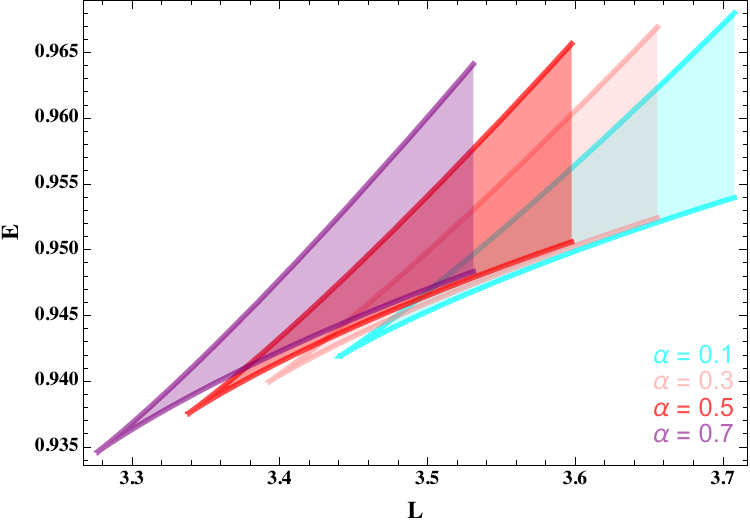} 
	\caption{\label{allowedLEregions} 
	The allowed $L-E$ parameter space for a timelike particle orbiting the RSBH at different values of $\alpha$.
	}
\end{figure}

\begin{figure}[htb]
	\centering 
	\includegraphics[width=.495\textwidth]{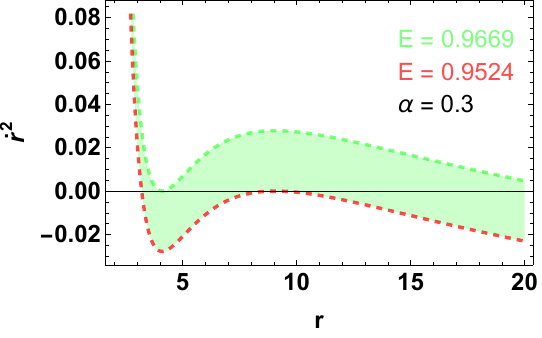} 
	\hfill
	\includegraphics[width=.495\textwidth]{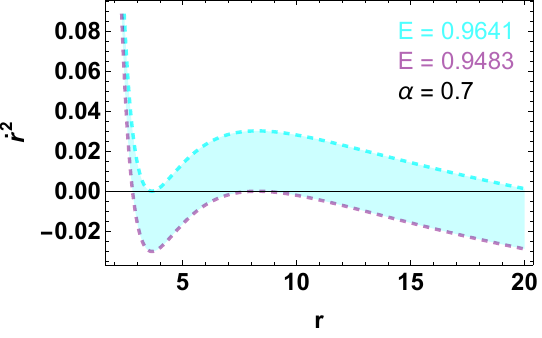} 
	\caption{\label{rdotsqu} 
		The radial motion $\dot{r}^2$ as a function of $r$ with $\alpha = 0.3$ (left panel) and $\alpha = 0.7$ (right panle) where $\epsilon = 0.5$.
		The shaded regions indicate the allowed energy ranges corresponding to bound orbits: $E \in (0.9524,\,0.9669)$ for $\alpha = 0.3$ (left panel) and $E \in (0.9483,\,0.9641)$ for $\alpha = 0.7$ (right panel). Only the orbits within these shaded regions, lying between the upper and lower energy bounds (dashed lines), represent bound orbits.
	}
\end{figure}

Figure \ref{veff} illustrates the behavior of the effective potential $V_{\text{eff}}$ as a function of the radial coordinate $r$ for $\alpha = 0.3$ and $\alpha = 0.7$, respectively. The sequence of curves (from bottom to top) shows that $V_{\text{eff}}$ increases with orbital angular momentum, ranging from $L_{\text{ISCO}}$ to $L_{\text{MBO}}$. The extrema of the effective potential, indicated by black dashed lines, correspond to stable and unstable circular orbits. As the orbital radius decreases, these extrema shift accordingly. Furthermore, the maximum value of $V_{\text{eff}}$ decreases with increasing $\alpha$. Finally, we note that the asymptotic behavior of the effective potential Eq. \eqref{effpot} is $V_{\text{eff}} \to 1$ as $r \to \infty$.
Using Eq. \eqref{LEcondition}, we plot in Fig. \ref{allowedLEregions} the allowed parameter space of orbital angular momentum and energy for bound orbits around the RSBH at different values of $\alpha$. The figure shows that, for a fixed orbital angular momentum, increasing $\alpha$ shifts the upper energy boundary of the bound orbits to higher values.
We now analyze the behavior of $\dot{r}^2$ for particles with parameters both within and beyond the allowed $L-E$ parameter space. Representative plots are shown in Fig. \ref{rdotsqu} for $\alpha = 0.3$ and $\alpha = 0.7$, with the angular momentum chosen as a linear combination of $L_{\rm ISCO}$ and $L_{\rm MBO}$, namely
\begin{equation}\label{angmom}
	L=L_{\mathrm{ISCO}}+\epsilon(L_{\mathrm{MBO}}-L_{\mathrm{ISCO}}),
\end{equation}
for different values of the energy.
The parameter $\epsilon$ is confined to the interval $(0,1)$, with $\epsilon=0$ representing the orbital angular momentum at the ISCO and $\epsilon=1$ corresponding to the MBO.  For $\epsilon > 1$, bounded orbits cease to exist.  Thus, by designating various values to $\epsilon$, the orbital angular momentum can be delineated. 

For $\epsilon = 0.5$, bound orbits exist only within the ranges $0.9524 \leq E \leq 0.9669$ for $\alpha = 0.3$ and $0.9483 \leq E \leq 0.9641$ for $\alpha = 0.7$, corresponding to cases where $\dot{r}^2 = 0$ admits at least two roots. For the upper dashed curve ($E = 0.9669$ for $\alpha = 0.3$, $E = 0.9641$ for $\alpha = 0.7$), bound orbits occur in the interval between the two roots. As the energy decreases, three roots appear, and the bound orbit lies between the last two roots in the region where $\dot{r}^2 > 0$. The bound orbit approaches the ISCO as the energy reaches $E = 0.9524$ ($E = 0.9483$), corresponding to the lower dashed curve.

For energies larger than these upper bounds ($E > 0.9669$ or $E > 0.9641$) or smaller than the lower bounds ($E < 0.9524$ or $E < 0.9483$), the equation $\dot{r}^2 = 0$ has either no root or only a single root, implying the absence of bound orbits. This behavior is illustrated outside the shaded regions in Fig. \ref{rdotsqu}.

\subsection{Null geodesic: Photon sphere, shadow silhouette and EHT constraints}\label{Sec3-shadow}

This section begins with an analysis of the photon sphere and shadow radius of the RSBH. 
The BH shadow boundary, as perceived by a far-off observer, delineates the apparent image of the photon region by discriminating between capture orbits and scattering orbits. The photon region actually represents the outermost limit of the spacetime zone associated with photon trajectories. In a spherically symmetric case, this corresponds to what is known as the photon sphere.
Considering the static spherically symmetric BH solution in \eqref{metric}, the Lagrangian \eqref{Lagrangian}
helps one to explain the motion of test particles.
Indeed, the equation $2\mathcal{L}(x,\dot{x}) = 0$, which describes the null geodesics, could be employed to describe the motion of photons.
To do so, it is sufficient to examine null geodesics within the equatorial plane $(\theta = \pi/2)$. 
Therefore, the null geodesic Lagrangian 
\begin{equation}\label{Lagrangian1}
	2\mathcal{L}(x,\dot{x}) =
	 -A(r)\dot{t}^{2}+\text{B}(r)^{-1}\dot{r}^{2} +r^{2}\sin^{2}\theta\dot{\varphi}^{2},
\end{equation}
provides the energy $E = A(r)dt/d\lambda$ and the angular momentum $L = r^{2} d\varphi/d\lambda$ as two conserved quantities, where the ratio $L/E$ is designated as the impact parameter $b=\frac{L}{E} = \frac{r^{2}}{A(r)} \frac{d\varphi}{dt}$, incorporating the coordinate angular velocity $d\varphi/dt$. The null geodesic $ds^{2} = 0$ allows us to establish a relationship between $r$ and $\varphi$ by employing the impact parameter $b$, as stated in \cite{PerlickPR2022}:
\begin{equation}\label{OrbitEq1}
	\left(\frac{dr}{d\varphi}\right)^{2} = r^{2}\text{B}(r)	\left(\frac{h(r)^{2}}{b^{2}}-1\right),
\end{equation}
with $h(r)^2 = \frac{r^{2}}{A(r)}$.
Since the orbit equation \eqref{OrbitEq1} relies only on the impact parameter $b$ at the trajectory's turning point $r = r_{ph}$, we must impose the condition $dr/d\varphi|_{r_{ph}} = 0$. This yields the relation $b^{-2} = A(r_{ph})/r_{ph}^{2}$ for the impact parameter at the turning point \cite{Chandrasekhar}. To derive the photon sphere radius $r_{ph}$, one must impose simultaneously the requirements $dr/d\varphi|_{r_{ph}} =0$ and $d^{2}r/d\varphi^{2}|_{r_{ph}} =0$, resulting in the following expression
\begin{equation}\label{rph1}
	\left.\frac{d}{dr}\left(\frac{r^{2}}{A(r)}\right)\right|_{r_{\text{ph}}} =0,
\end{equation}
which can be arranged as follows
\begin{equation}
r_{ph}A'(r_{ph})-2 A(r_{ph})=0.
\end{equation}
By numerically solving Eq. \eqref{rph1}, we investigate the influence of the deviation parameter $\alpha$ on the characteristics of the photon sphere radius $r_{ph}$. Figure \ref{sec2-fig2} shows a numerical plot that delineates the position of the photon sphere radius with respect to $\alpha$ values. This is significant as the critical impact parameter $b_{crit}$ can be obtained by $r_{ph}$, and both the shadow cast and the behavior of the shadow radius rely on the photon sphere radius.  Figure \ref{sec2-fig2} reveals that as the parameter $\alpha$ increases, both the photon sphere radius and the critical impact parameter drop.
In the standard Schwarzschild metric, the radii are defined as $r_{ph} = 3M$ and $b_{crit} = 3\sqrt{3}M$, respectively.

To create the BH shadow, we take all light rays coming from the static observer's position at $(t_{o},r_{o},\theta_{o} = \pi/2,\varphi_{o}=0)$ directed into the past. The observer is positioned at an angle $\theta_{sh}$ relative to the radial line that fulfills \cite{PerlickPRD2018}
 \begin{equation}
	\tan \theta_{sh} =\left. \lim_{\Delta x \to 0} \frac{\Delta y}{\Delta x} = \sqrt{A(r) r^{2}}\,\,\frac{d\varphi}{dr}\right|_{r=r_{\text{o}}}.
\end{equation}
Using basic trigonometry and representing $dr/d\varphi$ by the orbit equation \eqref{OrbitEq1}, we get
\begin{equation}
	\sin \theta_{sh} = \frac{b_{crit}}{r_{o}}\sqrt{A(r_{o})}.
\end{equation}
Thus, for a static observer located at a distance $r_{o}$ , the BH shadow radius becomes
 \begin{equation}\label{shadowradius}
	R_{sh} =  b_{crit}\sqrt{A(r_{o})}, \quad \text{or} \quad R_{sh} = r_{ph} \sqrt{\frac{A(r_{o})}{A(r_{ph})}}.
\end{equation}
The shadow radius manifestly relies on the observer's position, as seen by Eq.  \eqref{shadowradius}. It is observed that for an observer far from a BH with an asymptotically flat metric, it simplifies to $R_{sh}= b_{crit}$. The rationale for this simplification is because $A(r_{o}) \approx 1$ at a considerable distance from the BH \cite{VagnozziCQG2023}. This condition is satisfied for the Schwarzschild metric if $M \ll r_{o}$, indicating that the distance between the observer and the BH is substantially greater than its gravitational radius. Upon comparing the distances of M87* and Sgr A* from Earth, approximately $16.8 \,\text{Mpc}$ and $\approx 8 \text{kpc}$, respectively, to their gravitational radii, $r_{g} = \mathcal{O}(10^{-7})$ pc, it is obvious that this condition is fulfilled in both cases.

\begin{figure}[htb]
	\centering 
	\includegraphics[width=.5\textwidth]{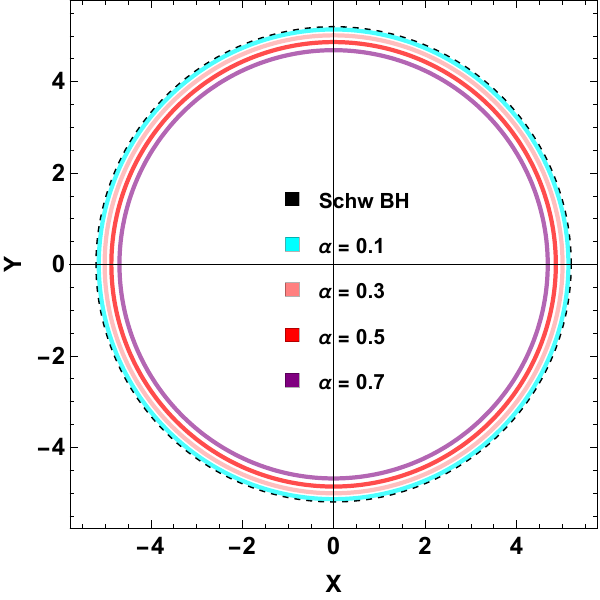} 
	\caption{\label{sec3-fig1} Observation of the RSBH shadow silhouettes as seen by an observer at spatial infinity, illustrating how the shadow varies with different values of the model parameter $\alpha$.}
\end{figure}

We are now in a position to explore the impact of the deviation parameter $\alpha$ on the shadow radius size for 
the spherically symmetric and static RSBH solutions
as observed by a distant observer at spatial infinity. Figure \ref{sec3-fig1} shows the shadow for different values of $\alpha$, related to the lapse function in Eq. \eqref{MetricFuncA1}.  The presence of the exponential suppression term in the RSBH metric, used in the literature to model quantum gravity or phenomenologically inspired regular BHs clearly influences the shadow size. 
In particular, as $\alpha$ grows with $p$ held constant, the shadow size decreases; whereas, the shadow size expands as $p$ increases with $\alpha$ held constant. Moreover, certain choices of $\alpha$ and $p$ indicated in Fig. \ref{sec3-fig1} lie in certain uncertainty bounds; so that the corresponding permissible ranges will be found using observational data from M87* and Sgr A*, as displayed below.

\begin{figure}[htb]
\centering 
\includegraphics[width=.32\textwidth]{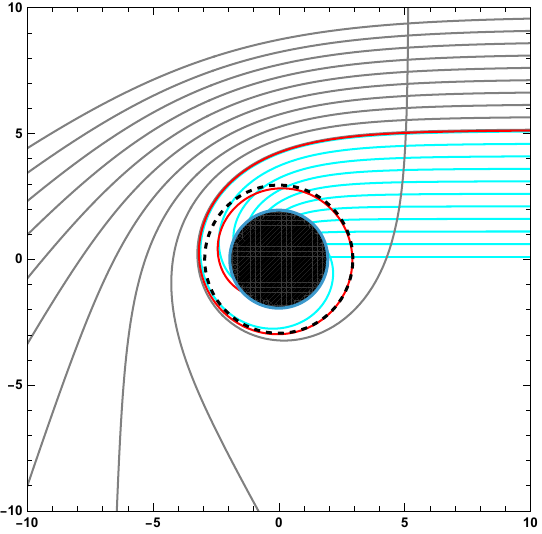} 
		\hfill
\includegraphics[width=.32\textwidth]{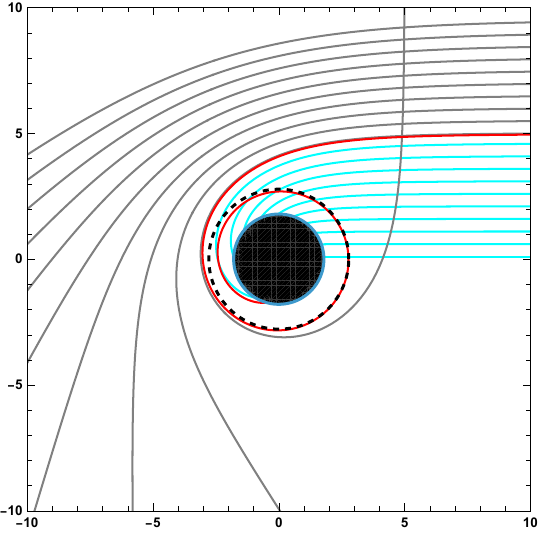}
		\hfill
\includegraphics[width=.32\textwidth]{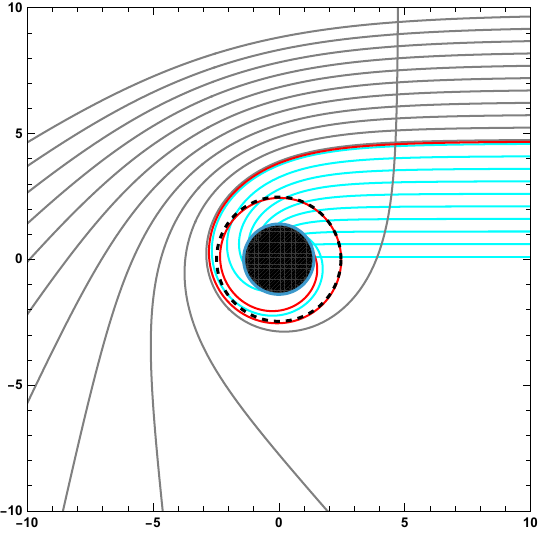} 
	\caption{\label{sec3-fig2} 
    Trajectories of light rays in polar coordinates $(r, \varphi)$, shown for different values of the deviation parameter $\alpha = 0.1, 0.35, 0.7$ (from left to right). The plots illustrate how variations in $\alpha$ influence the bending and propagation of light in the RSBH spacetime.}
\end{figure}

We then concentrate on the trajectory of a light ray traveling close to RSBH under varying $\alpha$ values.  Making the transformation $u=1/r$ helps one to investigate the photon trajectories around the BH by rearranging the orbit equation \eqref{OrbitEq1} as
\begin{equation}\label{OrbitEq2}
	\left(\frac{du}{d\varphi}\right)^{2} = 	\frac{1}{b^{2}}-u^{2}\text{B}(u).
\end{equation}
For light beams with $b > b_{crit}$, the turning point $u_{t}$ along their trajectory must be carefully treated. The location of the turning point can be obtained by solving the equation 
\begin{equation}\label{TurningPoint}
\frac{1}{b^{2}} - u_{t}^{2} \, \text{B}(u_{t}) = 0.
\end{equation}
The trajectories of light rays are displayed in Fig. \ref{sec3-fig2} by means of a ray-tracing code and Eq. \eqref{OrbitEq2} \cite{LiEPJC2021,HuEPJC2022}. Here, the distance between consecutive impact parameters $b_{crit}$ is uniformly set to $1/2$. The corresponding BH is shown by the solid black disk; the photon sphere is indicated by the dashed black circle.
It can be seen that every panel comprises gray lines and cyan lines, which correspondingly match the light beams of $b > b_{crit}$ and $b < b_{crit}$. The photon sphere $b = b_{crit}$ is the black dashed line; the image of the event horizon surface is shown as a black disk.
The colored photon trajectories illustrate three different types of photons traveling in the RSBH spacetime. Light rays characterized by $b = b_{crit}$ approach the photon sphere asymptotically, orbiting the BH along an unstable circular orbit infinitely many times. While light rays with $b > b_{crit}$ encounter the potential barrier and, after reaching their turning point, are deflected back towards infinity. Meanwhile, those with $b < b_{crit}$ move in the inward direction, inevitably being swallowed by the BH. Such light rays are not detectable by the distant observer, resulting in the formation of a shadow in the observational sky. Furthermore, it is observed that an increase in $\alpha$ results in a reduction of light rays absorbed by the BH, indicating a corresponding decrease in the shadow size.

The recent finding of BH shadows by the EHT Collaboration provided a compelling opportunity to conduct a rigorous test of gravitational theory in strong and relativistic field regimes. Furthermore, applying the Schwarzschild deviation parameter $\delta$ could also be useful for putting constraints on the model parameters of a particular BH. 
We then explore whether the model parameter $\alpha$ can be constrained employing the shadow image observational data released by the EHT for M87* and Sgr A* as well as the uncertainty reported in Refs. \cite{KocherlakotaPRD2021,VagnozziCQG2023}. The aim is to determine whether $\alpha$ lies within an observationally acceptable range, narrower than the theoretical interval $\alpha \in [0,\, 2 M^{4/3}/e]$, thereby ensuring that Eq. \eqref{metric} describes a regular BH with sub-Planckian curvature and a Minkowskian core—potentially making it a viable candidate for astrophysical BHs.

As noted in Ref. \cite{AkiyamaL12019}, the mass of the M87* BH is \( M_{M87*} = (6.5 \pm 0.7) \times 10^9 \, \text{M}_{\odot} \), its angular shadow diameter is \( \theta_{M87*} = 42 \pm 3 \, \mu\text{as} \), and it is situated at a distance of \( D_{\text{M87*}} = 16.8 \pm 0.8 \, \text{Mpc} \) from us. Considering the deviations of the Schwarzschild shadow, \( \delta_{M87*} = -0.01 \pm 0.17 \), where the expression \( \frac{R_{sh}}{M} = 3\sqrt{3}(1 + \delta_{M87*}) \) determines the shadow radius levels, the shadow size of M87* is restricted to the interval [4.26, 6.03] at the \( 1\sigma \) confidence level (CL).

For Sgr A*, the EHT collaboration \cite{AkiyamaL122022} thus reports an angular shadow diameter of $\theta_{\text{Sgr A*}} = 48.7 \pm 7 \, \mu\text{as}$. The estimated distance from Earth to Sgr A* is reported as $D_{\text{Sgr A*}} = 8277 \pm 9 \pm 33 \, \text{pc} \, (\text{VLTI})$ and $ 7953 \pm 50 \pm 32 \, \text{pc} \, (\text{Keck})$. Furthermore, the BH mass is determined to be $M_{\text{Sgr A*}} = (4.297\pm 0.012 \pm 0.040)\times10^{6} \text{M}_\odot \, \text{(VLTI)}$, $ (3.951\pm 0.047) \times 10^{6}\text{M}_\odot \, \text{(Keck)}$, $(4.0^{1.1}_{-0.6})\times10^{6}\text{M}_\odot \, \text{(EHT)}$. 
According to observations from Keck and VLTI, the fractional deviation from the expected Schwarzschild shadow for Sgr A* is quantified as $ \delta_{\text{Sgr A*}} = -0.08^{+0.09}_{-0.09} \, (\text{VLTI}) $ and $ \delta_{\text{Sgr A*}} = -0.04^{+0.09}_{-0.10} \, (\text{Keck}) $. The average of these values provides an estimate of $ \delta_{\text{Sgr A*}} \simeq -0.060^{+0.065}_{-0.065} \, (\text{Avg}) $. The relationship $\frac{R_{\text{S}}}{M} = 3\sqrt{3}(1+\delta_{\text{Sgr A*}})$, which defines the shadow radius with respect to the fractional deviation, reveals that the shadow size of Sgr A* is confined to the interval $[4.55, 5.22]$ at the $1\sigma$ CL.
We aim to apply these derived constraints to restrict the deviation of our BH from the standard Schwarzschild solution, particularly by analyzing how its properties diverge from the Schwarzschild scenario.

\begin{figure}[htb]
	\centering 
	\includegraphics[width=.49\textwidth]{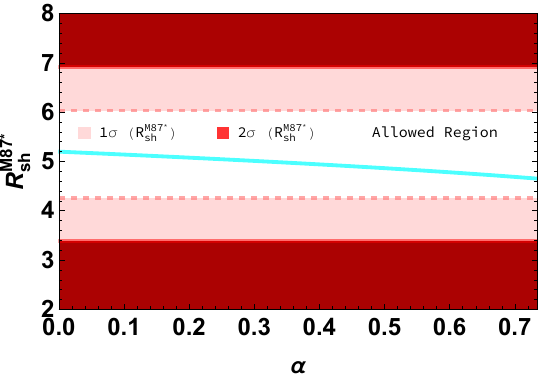}  
	\hfill
	\includegraphics[width=.50\textwidth]{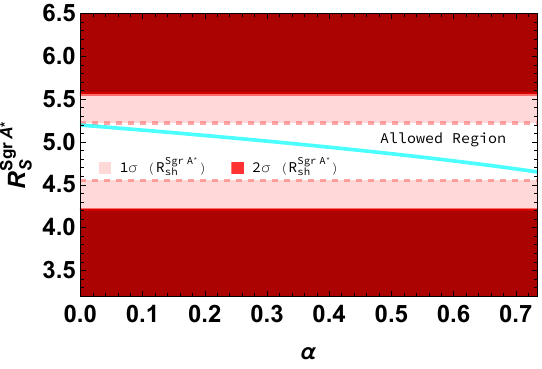} 
	\caption{\label{fig3-2}
	Shadow radius of the RSBH, derived from the metric function in Eq. \eqref{metric} and given in units of the BH mass M, plotted as a function of the parameter $\alpha$. The pink-shaded regions indicate values of $\alpha$ that are inconsistent with stellar dynamics observations for M87* (left panel) and Sgr A* (right panel). The white and pink areas correspond to the EHT horizon-scale images of M87* and Sgr A* at the $1\sigma$ and $2\sigma$ CLs, respectively. In the right panel, the shaded regions reflect values of \( \alpha \) that are consistent with the averaged Keck and VLTI mass-to-distance ratio priors for Sgr A*.
	}
\end{figure}

Figure~\ref{fig3-2} illustrates the variation of the shadow radius as a function of the parameter $\alpha$ for M87* and Sgr A*, including uncertainty at the $1\sigma$ and $2\sigma$ levels. 
As expected, an increase in $\alpha$ results in a reduction of the shadow radius of RSBH. We thus find that the stringent restrictions on the RSBH parameter $\alpha$ are observed at $p=1$. 

To determine the possible permitted ranges of the RSBH parameter $\alpha$, using both $1\sigma$ and $2\sigma$ CLs that were recorded by ETH for M87* (left panel) and Sgr A* (right panel), and using Eq. \eqref{shadowradius}, we see that the shadow radius behavior on $\alpha$ is a nonlinear behavior. It means that increasing $\alpha$ causes a decrease in the shadow radius, and they are proportional inversely to each other; therefore, it is necessary to plot this behavior to show such nonlinear behavior.

As shown in Fig. \ref{fig3-2}, the EHT observational data for M87* and Sgr A* indicate that, for $p = 2$, the parameter $\alpha$ is not subject to additional upper or lower bounds, since the shadow radius falls within the observationally allowed intervals. This implies that $\alpha$ need not be restricted to a narrower observational range than the theoretical interval $\alpha \in [0,\, 2 M^{4/3}/e]$. The model thus describes a regular black hole with sub-Planckian curvature and a Minkowskian core, making it a viable candidate for astrophysical black holes.

In addition, the shadow radius function in terms of $\alpha$ is a slow behavior function, especially for M87*. The shadow radius does not exceed the allowance zone of $1\sigma$ and $2\sigma$ CLs. 
We find that the RSBH parameter $\alpha$ is effective near the BH, as seen by its significant influence on the photon sphere radius. Therefore, even from a distance, the shadow cast can be changed. Thus, considering the results for $\alpha$, we get that M87* and Sgr A* BHs could be RSBHs with the current precision of astrophysical data.

\section{Periodic orbits near a regularized Schwarzschil black hole}\label{POs}
Following the examination of ISCOs and MBOs, we now concentrate on periodic orbits in the vicinity of the RSBH model. 
According to Ref. \cite{LevinPRD2008}, generic orbits can be considered as perturbations of periodic orbits.
 Studying periodic orbits is therefore particularly insightful, as it not only clarifies the structure of generic orbits but also deepens our understanding of the associated gravitational radiation.
It is rather significant for the investigation of gravitational wave radiation.
These kinds of bound orbits can be distinguished by their repetitive trajectories. 
The background is that of a spherically symmetric BH, with $\theta = \pi/2$ and $\dot{\theta} = 0$, hence
periodicity is observed when the frequency ratio of the radial $(\omega_r)$ to the azimuthal $(\omega_\varphi)$ oscillations is a rational number. 
In this way, we apply the classification scheme proposed in Ref. \cite{LevinPRD2008} to index various periodic orbits near the RSBH using a triplet of integers $(z; w; \nu)$, which correspond to the zoom, whirl, and vertex behaviors, respectively.
One should note that these three parameters are all integers.
Thus, one can define a rational number (or a frequency ratio) $q$ as the relationship between the two frequencies, $(\omega_r)$ and $(\omega_\varphi)$ of oscillations in the radial $r$-motion and the azimuthal $\varphi$-motion, in terms of three integers $(z; w; \nu)$ as
\begin{equation}\label{rationalnumber}
q=\frac{\omega_{\varphi}}{\omega_{r}}-1=\frac{\Delta\varphi}{2\pi}-1=w+\frac{\nu}{z}.
\end{equation}
The ratio of azimuthal to radial frequencies is defined by $\frac{\omega_{\varphi}}{\omega_{r}} = \frac{\Delta \varphi}{2\pi}$, where $\Delta\varphi \equiv \oint d\varphi$ is the total azimuthal angle swept during one radial period, which must equal an integer multiple of $2\pi$. Employing the geodesic equations of the RSBH spacetime, the quantity $q$ can then be expressed as:
\begin{equation}\label{rnq}
\begin{split}
q &= \frac{1}{\pi} \int_{r_1}^{r_2}\,\frac{\dot{\phi}}{\dot{r}}dr-1\\
& = \frac{1}{\pi} \int_{r_1}^{r_2} \frac{L}{r^2} \frac{1}{\sqrt{E^2-V_{\rm eff}}} \,dr -1,
\end{split}
\end{equation}
where $r_{1}$ and $r_{2}$ are referred to as the radius of periapsis and apoapsis of the periodic orbits, respectively. 
These two turning points of the particle motion are obtained by the equation of motion \eqref{rdot2}, namely the two roots of $\dot{r}^{2}=0$.
For $q$ as an irrational number, the orbit of a timelike particle exhibits precessional motion, analogous to the perihelion precession of Mercury. In contrast, if $q$ takes a rational number, the particle’s trajectory becomes closed, giving rise to a periodic orbit.
The parameter $q$ quantifies the degree of periapsis precession beyond that of a closed elliptical orbit, offering valuable insight into the geometric structure of the trajectory. Furthermore, the classification scheme also accounts for the  leaf-tracing order, which denotes the sequence in which the orbital segments (or "leaves") are traversed. Together, these elements provide a detailed framework for understanding the intricate dynamics of periodic orbits in spacetime. 
So, studying periodic orbits gives us important information about how generic orbits behave and the gravitational waves produced near BHs.

The apsidal angle $\Delta\varphi$ denotes the azimuthal angular variation over each period and is expressed as:
\begin{equation}\label{dphi}
\begin{split}
\Delta\varphi =\oint\, \, d\phi = 2\int_{\phi_1}^{\phi_2}\,d\phi & = 2 \int_{r_1}^{r_2}\,\frac{\dot{\phi}}{\dot{r}}dr \\
&= 2 \int_{r_1}^{r_2} \frac{L}{r^2} \frac{1}{\sqrt{E^2-V_{\rm eff}}} \,dr.
\end{split}
\end{equation}
The symmetrical characteristics of the trajectory, taken by the timelike particle, lead to a coefficient of $2$.  The apsidal angle $\Delta\varphi$ spanned by a particle is not solely governed by its energy and angular momentum -- it is also described by the surrounding spacetime geometry, as reflected in the function $A(r)$. Black holes with various parameter values will thus show distinct apsidal angles.

\begin{figure*}
	\centering
	\includegraphics[width=7.33 cm]{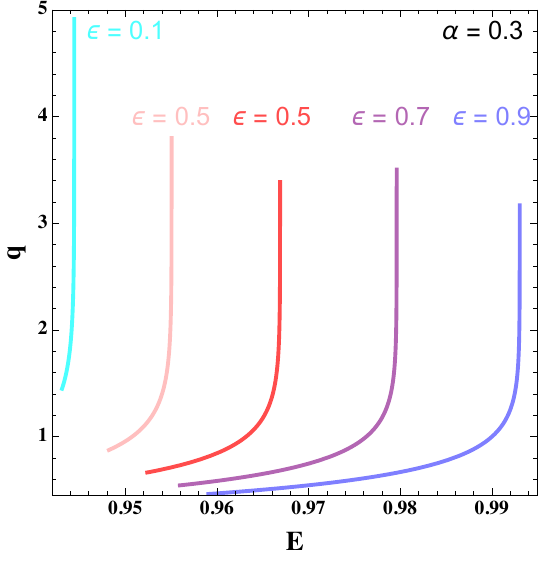}
	\includegraphics[width=7.67 cm]{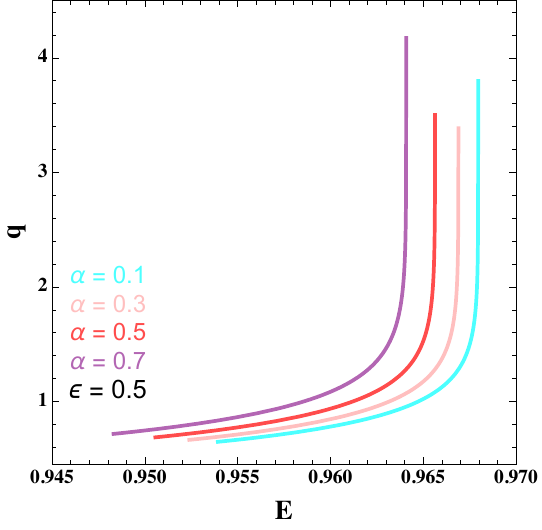}

	\vspace{1pt} 
	
	\includegraphics[width=7.33 cm]{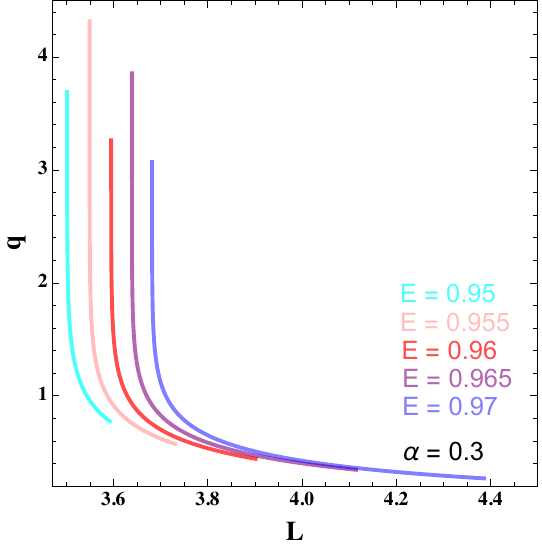}
	\includegraphics[width=7.67 cm]{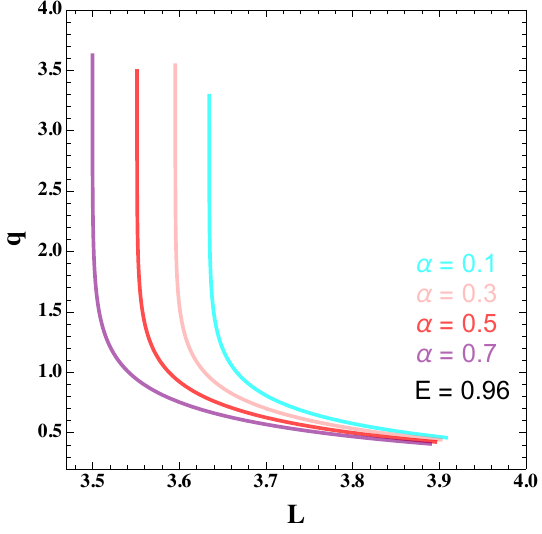}
	
	\captionsetup{justification=raggedright,singlelinecheck=false}
	\caption{The rational number $q$ as a function of the energy  (top panels) and angular momentum  (bottom panels) of periodic orbits around RSBHs}
	\label{RatNamq}
\end{figure*}

Figure \ref{RatNamq} illustrates the behavior of the rational number $q$ as a function of the energy $E$ and angular momentum $L$. The top panels show the dependence of $q$ on the particle energy. We find that $q$ increases slowly with $E$ at first, but then rises sharply as the energy approaches its maximum value. In the upper left panel, the variation of $q$ with different values of $\epsilon$ is displayed for the case $\alpha = 0.3$, revealing that $\epsilon$ strongly influences the behavior of $q$ as a function of $E$. In the upper right panel, we instead fix $\epsilon = 0.5$ and examine how different choices of $\alpha$ affect $q$.
The lower panels present the behavior of $q$ as a function of angular momentum $L$. In both cases -- whether varying $E$ (lower left panel) or $\alpha$ (lower right panel) -- $q$ decreases gradually with increasing $L$, reaching its minimum value at the lowest allowed angular momentum. Furthermore, Fig. \ref{RatNamq} shows that, for a fixed rational number $q$, both the energy and angular momentum of the periodic orbits decrease as the parameter $\alpha$ increases.
We concentrate on specific periodic orbits with comparatively low values of $q$ to exemplify the periodic orbits and associated GW radiations for two primary reasons. I) large-$q$ orbits correspond to extreme “whirl-dominated” trajectories near the ISCO. Such orbits are typically short-lived, as they dissipate energy rapidly through GW radiation, and are therefore unlikely to be of astrophysical relevance. II) large-$q$ orbits require finely tuned initial conditions and are extremely sensitive to perturbations, including those induced by gravitational wave back-reaction. This sensitivity is clearly reflected in Fig. \ref{RatNamq}, where even small variations in energy or angular momentum cause sharp increases in $q$.

For periodic orbits labeled by different $(z; w; \nu)$ around the regular BH with sub-Planckian curvature and a Minkowskian core, and for various values of the model parameter $\alpha$, we numerically determine the orbital angular momentum $L$ of the periodic orbits while keeping the energy fixed at $E = 0.96$. Conversely, we compute the energy $E$ of the periodic orbits at fixed angular momentum $L$, obtained by setting $\epsilon = 0.5$ in Eq. \ref{angmom}.

\begin{figure*}
	\centering
	\includegraphics[width=5.0cm]{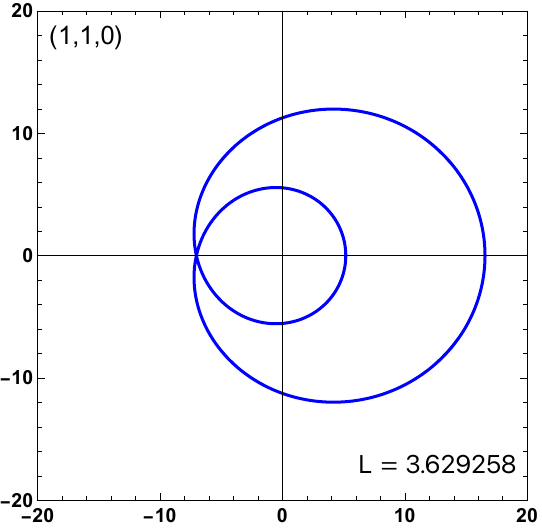} 
	\includegraphics[width=5.0cm]{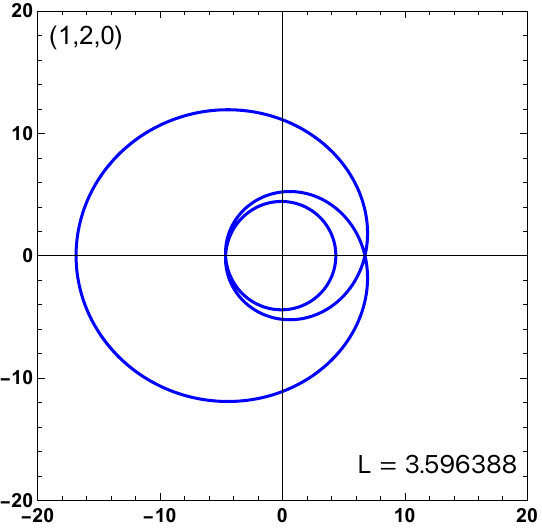}
	\includegraphics[width=5.09cm]{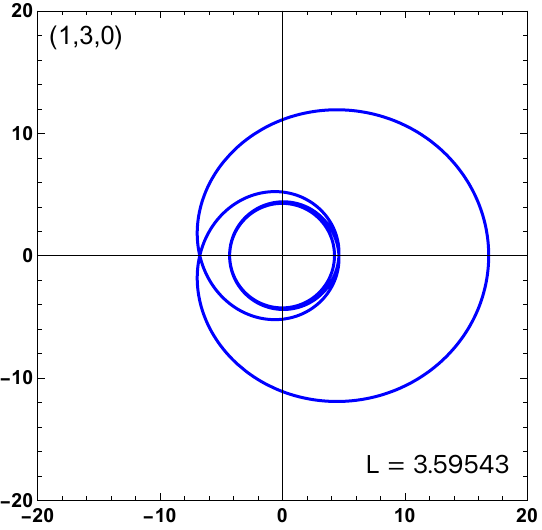} 
	\includegraphics[width=5.0cm]{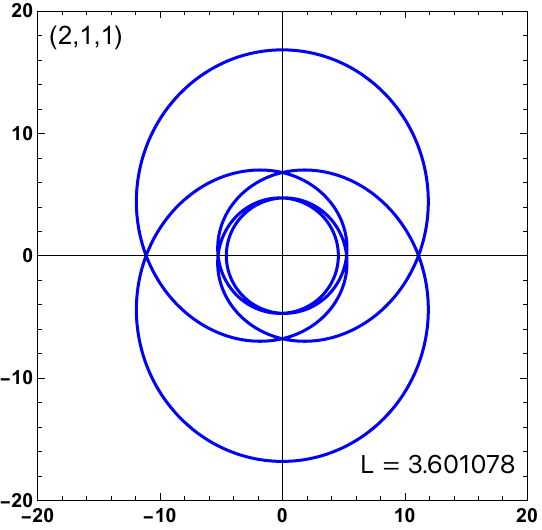} 
	\includegraphics[width=5.0cm]{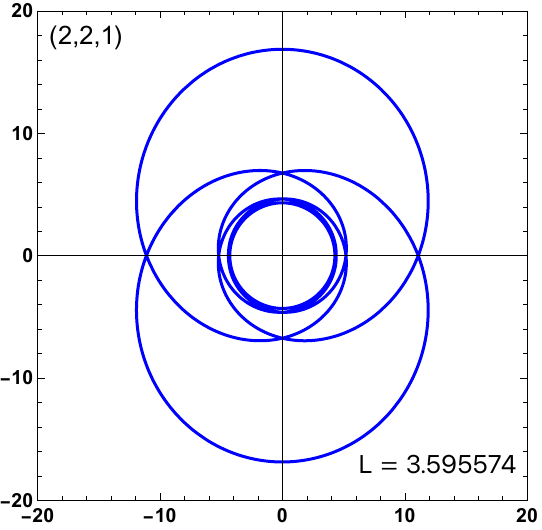} 
	\includegraphics[width=5.0cm]{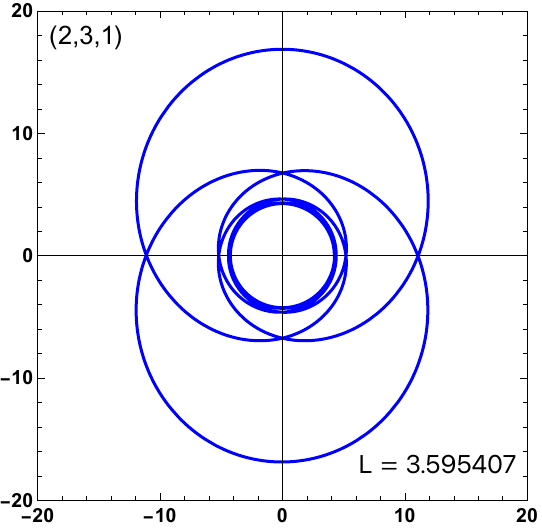} 
	\includegraphics[width=5.0cm]{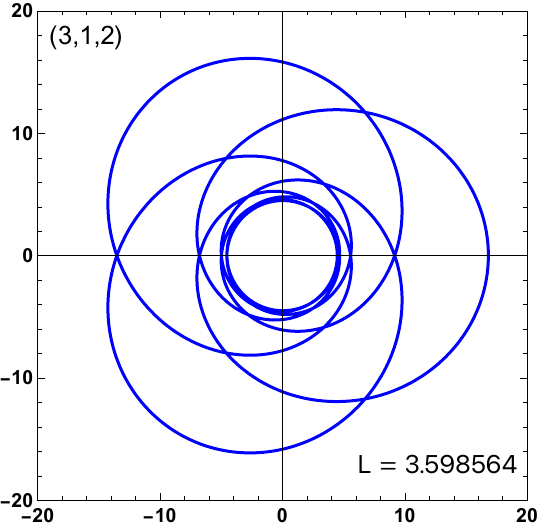} 
	\includegraphics[width=5.0cm]{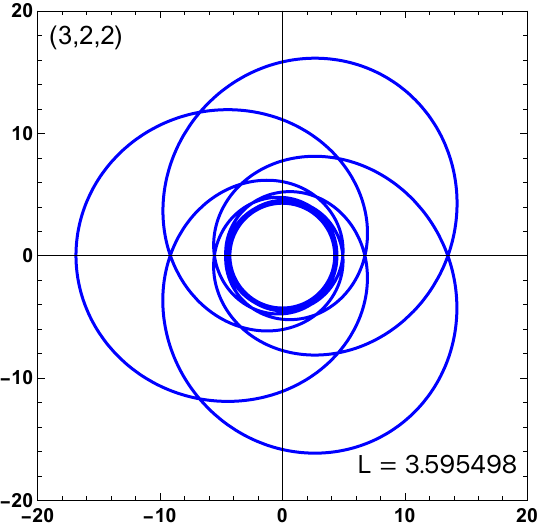}
	\includegraphics[width=5.0cm]{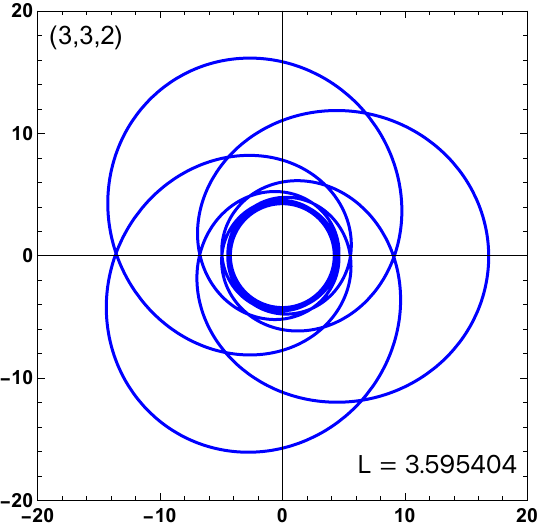} 
	\includegraphics[width=5.0cm]{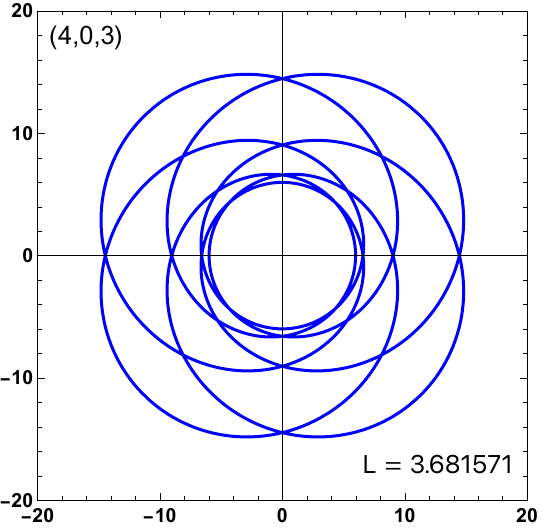} 
	\includegraphics[width=5.0cm]{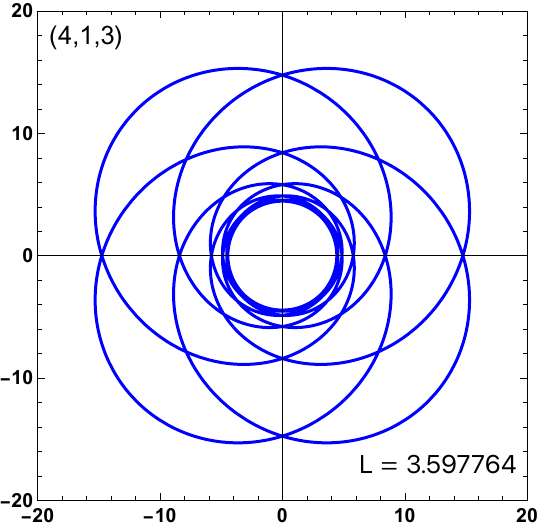} 
	\includegraphics[width=5.0cm]{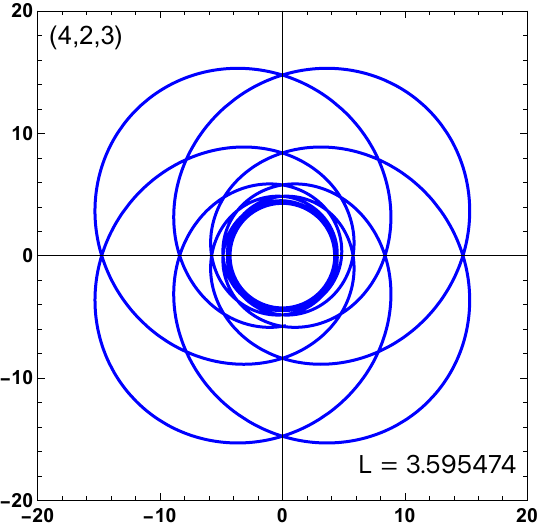} 
	\captionsetup{justification=raggedright,singlelinecheck=false}
	\caption{Periodic orbits for various values of $(z; w; \nu)$ around the RSBH. The value of model parameter is $\alpha=0.3$ and $E=0.96$.}
	\label{periodic1}
\end{figure*}

\begin{figure*}
	\centering
	\includegraphics[width=5.0cm]{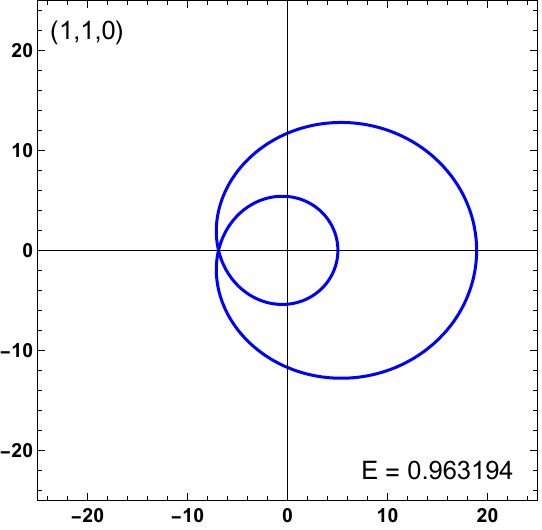} 
	\includegraphics[width=5.0cm]{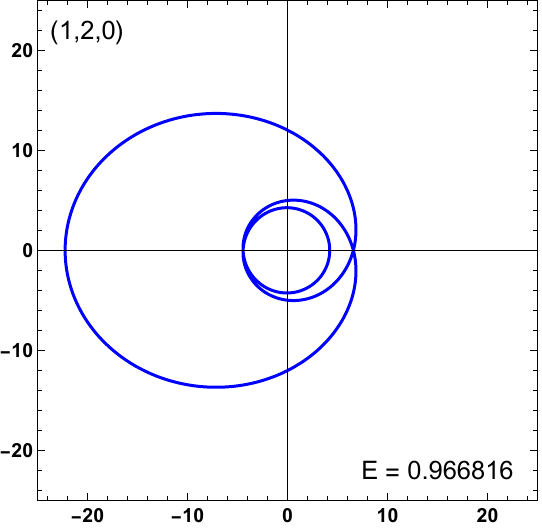}
	\includegraphics[width=5.0cm]{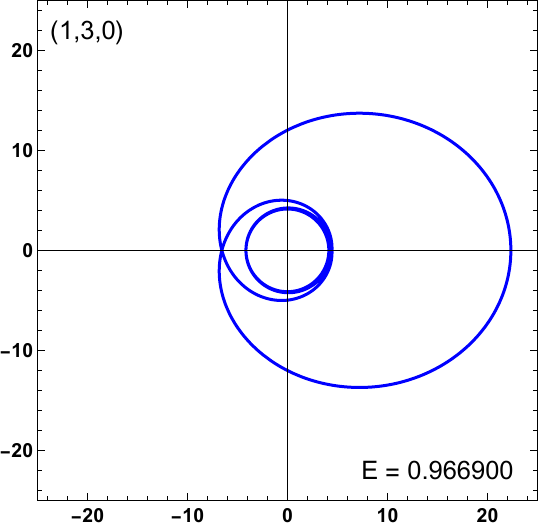} 
	\includegraphics[width=5.0cm]{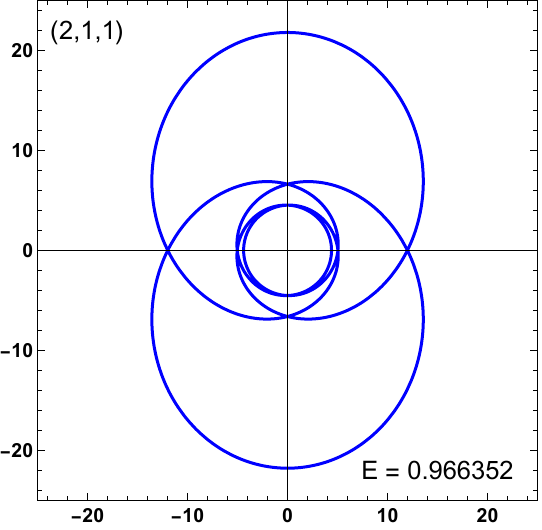} 
	\includegraphics[width=5.0cm]{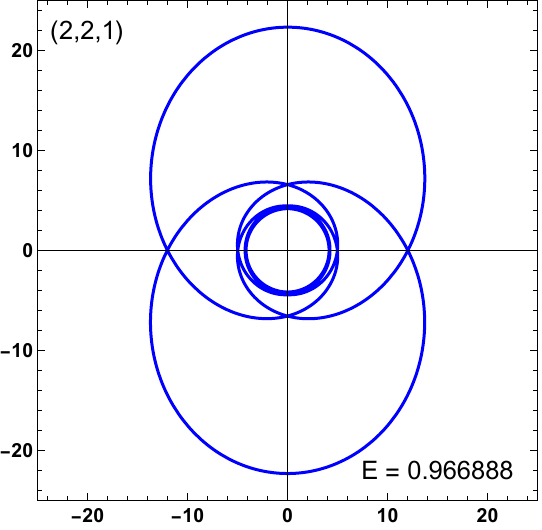} 
	\includegraphics[width=5.0cm]{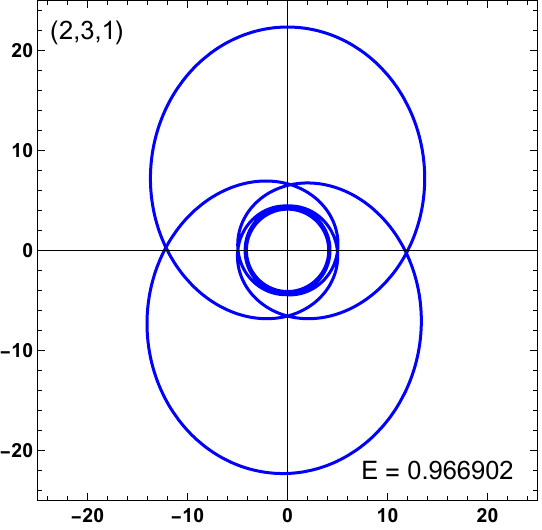} 
	\includegraphics[width=5.0cm]{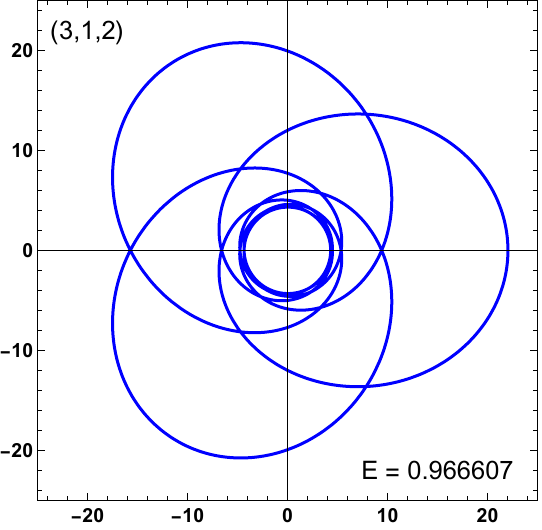} 
	\includegraphics[width=5.0cm]{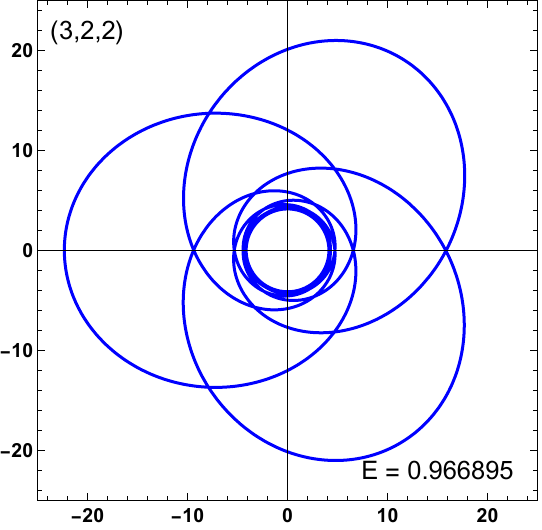} 
	\includegraphics[width=5.0cm]{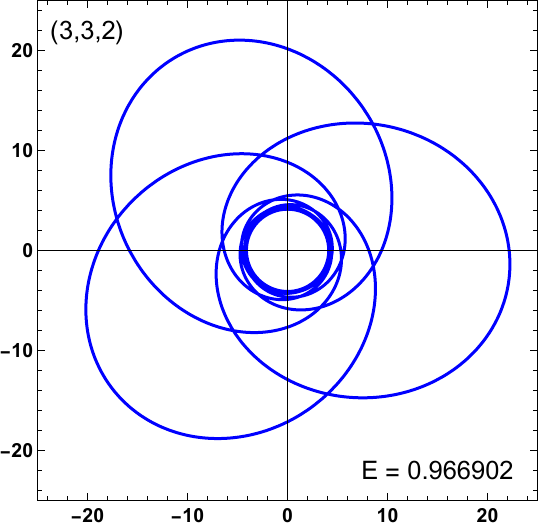}
	\includegraphics[width=5.0cm]{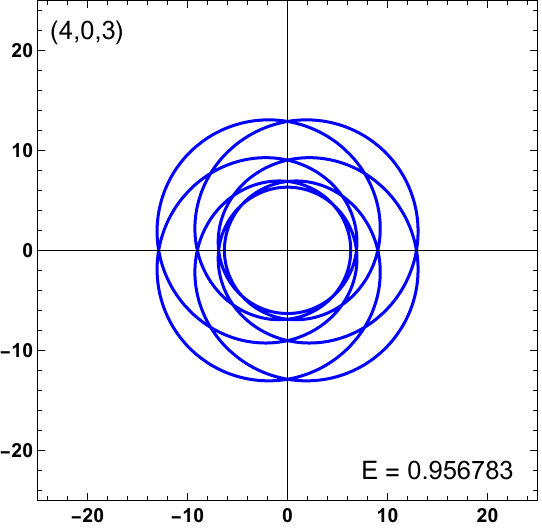} 
	\includegraphics[width=5.0cm]{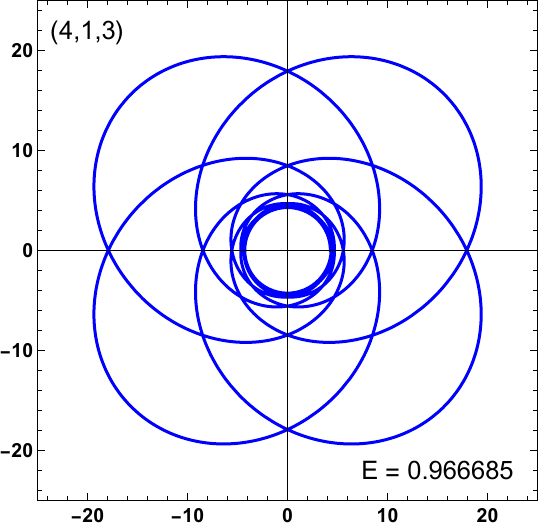} 
	\includegraphics[width=5.0cm]{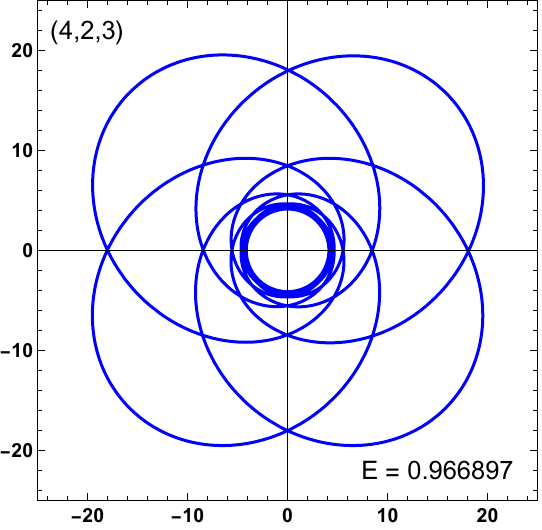} 
	\captionsetup{justification=raggedright,singlelinecheck=false}
	\caption{Periodic orbits for various values of $(z; w; \nu)$ around the RSBH. The value of th model parameter is $\alpha=0.3$ and $\epsilon=0.5$.}
	\label{periodic2}
\end{figure*}

Figure \ref{periodic1} displays the trajectories of periodic orbits characterized by different $(z; w; \nu)$ in the $r- \varphi$ plane for a fixed parameter $\alpha = 0.3$ and particle energy $E = 0.96$, in the background of a regular black hole with sub-Planckian curvature. The horizontal and vertical axes correspond to $r \cos \varphi$ and $r \sin \varphi$, respectively, while the values of $q$ and $E$ associated with each orbit are indicated in the subfigures. Figure \ref{periodic2} presents the corresponding trajectories for various $(z; w; \nu)$ with $\alpha = 0.3$ and $\epsilon = 0.5$ ($L = 3.65576$). These plots demonstrate that the integer $z$ specifies the number of leaf-like structures in the trajectory: as $z$ grows, the leaves enlarge and the orbital patterns become increasingly intricate.

Table \ref{Tab:EList} lists the particle energies associated with different $q$-periodic orbits for various values of the model parameter $\alpha$, with $\epsilon = 0.5$ fixed. The corresponding orbital angular momenta are given in the second column. As the RSBH parameter increases, both the orbital angular momentum and particle energy decrease, reaching their maximum values in the Schwarzschild limit.

Table \ref{Tab:LList} presents the orbital angular momentum $L$ for periodic orbits labeled by $(z; w; \nu)$ at fixed energy $E = 0.96$ for various choices of $\alpha$. Consistent with the trend observed in Table 1, the orbital angular momentum decreases as the model parameter $\alpha$ increases.

\begin{table*}[htbp]
	\caption{ 
		The particle energy $E$ of the periodic orbits determined by various $(z; w; \nu)$ configurations and different model parameter $\alpha$, with $\epsilon = 0.5$.
		\label{Tab:EList}}
	\begin{center}
		\vspace{-0.2cm}
		\resizebox{0.999\textwidth}{!}{ 
		\begin{tabular}{ccccccccccc}
			\toprule[0.5pt]\toprule[0.5pt]
			$\alpha$  & L & $E_{(1;1;0)}$    & $E_{(1;2;0)}$  & $E_{(2;1;1)}$  &  $E_{(2;2;1)}$  & $E_{(3;1;2)}$  & $E_{(3;2;2)}$  & $E_{(4;1;3)}$ & $E_{(4;2;3)}$ \\
			\midrule[0.5pt]
			0.1  & 3.707803 &0.964753 & 0.967780 & 0.967513 & 0.967957 & 0.967727 & 0.967590 & 0.967793 & 0.967963 \\
			0.3   & 3.655757 &0.963193 & 0.966816 & 0.966351 & 0.966888 & 0.966606 & 0.966894 & 0.966685 & 0.966896 \\
			0.5   & 3.597712 &0.961228 & 0.965523 & 0.964941 & 0.965620 & 0.965256 & 0.965629 & 0.965355 & 0.965632 \\
			0.7   & 3.531208 &0.958585 & 0.953900 & 0.963131 & 0.964055 & 0.963546 & 0.964069 & 0.963680 & 0.964074 \\
			\bottomrule[0.5pt] \bottomrule[0.5pt]
		\end{tabular}
	}
	\end{center}
\end{table*}

\begin{table*}[htbp]
	\caption{The angular momentum $L$ of the periodic orbits determined by various $(z; w; \nu)$ configurations and different model parameter $\alpha$ with energy $E = 0.96$. \label{Tab:LList}}
	\begin{center}
		\vspace{-0.2cm}
		\resizebox{0.999\textwidth}{!}{ 
		\begin{tabular}{cccccccccc}
			\toprule[0.5pt]\toprule[0.5pt]
			$\alpha$  & $L_{(1;1;0)}$    & $L_{(1;2;0)}$  & $L_{(2;1;1)}$  &  $L_{(2;2;1)}$  & $L_{(3;1;2)}$  & $L_{(3;2;2)}$  & $L_{(4;1;3)}$ & $L_{(4;2;3)}$ \\
			\midrule[0.5pt]
			0.1   & 3.666319 & 3.635396 & 3.639719 & 3.634664 & 3.637390 & 3.634596 & 3.636654 & 3.634575 \\
			0.3   & 3.629258 & 3.596388 & 3.601078 & 3.595574 & 3.598564 & 3.595498 & 3.597764 & 3.595474 \\
			0.5   & 3.588098 & 3.552351 & 3.557624 & 3.551400 & 3.554820 & 3.551308 & 3.553919 & 3.551278 \\
			0.7   & 3.541551 & 3.501282 & 3.507554 & 3.500072 & 3.504265 & 3.499949 & 3.503190 & 3.499908  \\
			\bottomrule[0.5pt] \bottomrule[0.5pt]
		\end{tabular}
	}
	\end{center}
\end{table*}

\section{Numerical Kludge gravitational wave radiation from periodic orbits}\label{GWs}

In an EMRI system, a regular BH with sub-Planckian curvature and a Minkowskian core, regarded as a type of RSBH, functions as a supermassive BH; thus, modeling the RSBHs as M87* and Sgr A* in subsection \ref{Sec3-shadow}, we show that they can be candidates for supermassive BHs. 
A small-mass compact object moving periodically in the strong gravitational field of a RSBH can emit GWs. Given that the mass of the orbiting body is significantly smaller than that of the RSBH, its influence on the background spacetime is negligible and can be treated as a perturbation.
In this scenario, it is logical to use the adiabatic approximation \cite{HughesPRD2000,HughesPRD2001,GlampedakisPRD2002,HughesPRL2005,DrascoCQG2005,GairPRD2005,GlampedakisCQG2005,DrascoPRD2005,SundararajanPRD2007,SundararajanPRD2008,MillerPRD2021,IsoyamaPRL2022}, that is, the energy and angular momentum losses of the system can be disregarded inside one or many orbital periods.  We suppose, under the adiabatic approximation, that the system's angular momentum and energy stay unaltered inside one period.
Thus, one can consider these conserved quantities, that is, energy and angular momentum, as fixed values to compute the GW radiation in one period. In EMRI systems, this treatment approach is efficacious. It can quickly and precisely manifest the orbital evolution of a minor celestial object in the RSBH background, together with its GW properties. This approach has been widely used in references \cite{ZhaoEPJC2024,QiEPJC2024,YangJCAP2025,TuPRD2023,ZiPLB2024,ZiPRD2024}.

We derive the GW signal using the kludge waveform method presented in Ref. \cite{BabakPRD2007}. 
In this work, the method proceeds in two main steps. I) the periodic orbital motion of the small object is obtained by numerically solving the timelike geodesic equations in the RSBH spacetime. II) After addressing the orbital motion, we use the quadrupole formula to find the gravitational waveform. 
This approach provides a preliminary exploration of the GW signals from EMRIs and their potential to probe both the orbit and the central RSBH model.

The small-mass object travels in a distinct bound trajectory, oscillating between the orbit's apoapsis and periapsis, as it travels in a periodic orbit around the supermassive BH.
Notably, the particle undergoes long-duration, high-frequency oscillations close to the event horizon as it spirals inward, producing strong gravitational wave signals. 
These waves transmit key information regarding the physics close to the BH's event horizon, offering them one of the most promising instruments for investigating the BH.
The GWs produced by EMRIs provide a direct connection between the timelike particle motion and the BH physics. This is a crucial means to investigate both the dynamics of compact objects and the BH properties.

The selection of a coordinate system is essential for the computation and analysis of gravitational waves.  The geodesic equations are typically solved in Boyer-Lindquist coordinates $(r; \theta; \varphi)$, but the waveform is often expressed in a detector-adapted coordinate system $(X;Y;Z)$.  This transformation streamlines the examination of the signal recorded by a gravitational wave detector.  The explicit transformation from Boyer–Lindquist to Cartesian coordinates is presented in \cite{BabakPRD2007}:
\begin{equation}\label{cartesian coor}
	x=r \sin \theta \cos \phi, \quad y=r \sin \theta \sin \phi, \quad z=r \cos \theta .
\end{equation}

Once the particle’s orbit is constructed in the pseudo-flat background spacetime, we proceed by applying the wave-generation formula in flat spacetime. 
Within the weak-field, linearized approximation, the spacetime metric is written as
$g_{\mu\nu} = \eta_{\mu\nu} + h_{\mu\nu}, \,\text{and}\,  
\bar{h}^{\mu\nu} \equiv h^{\mu\nu} - \tfrac{1}{2} h \, \eta^{\mu\nu}$,
where $\eta_{\mu\nu}$ denotes the Minkowski metric, and $h_{\mu\nu}$ a small perturbation satisfying $|h_{\mu\nu}| \ll 1$ with trace $h = \eta^{\mu\nu} h_{\mu\nu}$. Imposing the Lorentz gauge condition, $\bar{h}^{\mu\alpha}{}_{,\alpha} = 0$, the linearized Einstein field equation is given by
\begin{equation}\label{linearizedEFE}
	\square\bar{h}^{\mu\nu}=-16\pi\mathcal{T}^{\mu\nu}.
\end{equation}
Here, $\square$ denotes the flat-spacetime d’Alembertian operator, while the effective energy–momentum tensor $\mathcal{T}^{\mu\nu}$ obeys the conservation law $\mathcal{T}^{\mu\nu}{}_{,\nu} = 0$. In a coordinate system centered on the BH, a particular solution to this equation is provided by the retarded potential:

\begin{equation}\label{retardedpotential}
	h_{\mu\nu}(\mathbf{x},t)=4G \int d^3\mathbf{x}^{\prime} \frac{\mathcal{T}^{\mu\nu}(\mathbf{x}^{\prime},t-|\mathbf{x}-\mathbf{x}^{\prime}|)}{|\mathbf{x}-\mathbf{x}^{\prime}|},
\end{equation}
where, the coordinates of the observer and source are denoted by $\mathbf{x}^{\prime}$ and $\mathbf{x}$, respectively. This expression describes the gravitational radiation generated by the source term $\mathcal{T}^{\mu\nu}$. The observer coordinate $\mathbf{x}^{\prime}$ acts as an integration variable spanning the spatial region where the effective energy–momentum tensor $\mathcal{T}^{\mu\nu}$ is non-vanishing. When the motion of the source is only weakly affected by gravity, $\mathcal{T}^{\mu\nu}$ may be approximated by the matter energy–momentum tensor $T^{\mu\nu}$ \cite{BabakPRD2007}. In the slow-motion limit, the Press formula then reduces to the traditional quadrupole formula \cite{BabakPRD2007}

\begin{equation}\label{quadrupoleformula}
	\bar{h}^{i j}(t, \mathbf{x})=\frac{2}{r}\left[\ddot{I}^{i j}\left(t^{\prime}\right)\right]_{t^{\prime}=t-r},
\end{equation}
in which 
\begin{equation}\label{quadrupolemoment}
	I^{i j}=\int  x^i x^j T^{t t}\left(t, x^i\right) d^3x,
\end{equation}
stands for the mass quadrupole moment of the source. For a small-mass object following the trajectory $Z^i(t)$, the relevant component of the stress-energy tensor is the $T^{tt}$ term, which takes the form \cite{ThornePMP1980}:
\begin{equation}\label{stress-energy tensor}
	T^{t t}\left(t, x^i\right)=m \delta^3\left(x^i-Z^i(t)\right) .
\end{equation}

Upon inserting Eq. \eqref{quadrupolemoment} into Eq. \eqref{quadrupoleformula}, we arrive at the gravitational-wave quadrupole expression in the slow-rotation approximation:
\begin{equation}\label{GW}
	h_{ij}=\frac{2}{D_{\mathrm{L}}}\frac{d^2I_{ij}}{dt^2}=\frac{2m}{D_{\mathrm{L}}}(a_ix_j+a_jx_i+2v_iv_j),
\end{equation}

In this context, $D_{\mathrm{L}}$ stands for the luminosity distance from the EMRIs to the detector, whereas $v_i$ and $a_i$ signify the spatial velocity and acceleration of the small-mass object, respectively.
To investigate the GW signal as observed by a detector, we introduce a detector-adapted coordinate system $(X, Y, Z)$, centered on the BH. This frame is oriented with respect to the original $(x, y, z)$ system through the inclination angle $\iota$ and the longitude of pericenter $\zeta$. In terms of the original coordinates, the unit vectors of the detector frame are given by:

\begin{eqnarray}
	\hat{e}_X &=& (\cos\zeta, -\sin\zeta, 0),\\
	\hat{e}_Y &=& (\sin\iota \sin\zeta, \cos\iota \cos\zeta, -\sin\iota),\\
	\hat{e}_Z &=& (\sin\iota \sin\zeta, -\sin\iota \cos\zeta, \cos\iota),
\end{eqnarray}
The GW polarisations $h_+$ and $h_\times$ are derived by projecting $h_{ij}$, as described in Eq. \eqref{GW}, yielding:
\begin{eqnarray}\label{4.5}
	h_+&=\frac{1}{2}\big(e_X^i e_X^j-e_Y^ie_Y^j\big)h_{ij},\\
	h_{\times}&=\frac{1}{2}\big(e_X^i e_Y^j-e_Y^ie_X^j\big)h_{ij},
\end{eqnarray}
the above polarisations can be expressed using the components $h_{\zeta\zeta}$, $h_{\iota\iota}$, and $h_{\iota\zeta}$, identified in 
the detector frame as specific linear combinations of the $h_{ij}$ components as follows:

\begin{eqnarray}\label{4.64}
	h_+&=&\frac{1}{2}\big(h_{\zeta\zeta}-h_{\iota\iota}\big),\\\label{4.6}
	h_{\times}&=&h_{\iota\zeta},
\end{eqnarray}
in which the corresponding components are given by \cite{BabakPRD2007}
\begin{eqnarray}\label{4.7}
	h_{\zeta\zeta}&=&h_{xx}\cos^2\zeta-h_{xy}\sin{2 \zeta}+h_{yy}\sin^2\zeta,\\
	h_{\iota\iota}&=& \cos^2\iota\big[h_{xx}\sin^2\zeta + h_{xy}\sin 2 \zeta + h_{yy}\cos^2 \zeta\big] \nonumber \\
	&& +h_{zz} \sin^2\iota - \sin{2 \iota}\big[h_{xz \sin\zeta}+h_{yz}\cos\zeta\big],\\
	h_{\iota\zeta}&=& \frac{1}{2}\cos\iota\big[h_{xx} \sin {2 \zeta}+ 2 h_{xy}\cos{2 \zeta}- h_{yy}\sin{2 \zeta}\big]\nonumber\\
	&&+\sin\iota\big[h_{yz} \sin\zeta-h_{xx} \cos\zeta\big].
\end{eqnarray}

To illustrate the gravitational waveforms associated with various periodic orbits, and to examine the influence of the RSBH model parameter $\alpha$, we take an EMRI system composed of a small compact object of mass $m = 10 M_\odot$ orbiting a supermassive BH of mass $M = 10^6 M_\odot$, where $M_\odot$ denotes the solar mass. For simplicity, we fix the inclination angle at $\iota = \pi/4$, the longitude of pericenter at $\zeta = \pi/4$, and the luminosity distance at $D_L = 200 \, \mathrm{Mpc}$.

\begin{figure*}
	\centering
	\begin{tabular}{c@{\hspace{15pt}}c}
		\begin{minipage}{0.4\textwidth}
			\centering
			\includegraphics[width=\linewidth]{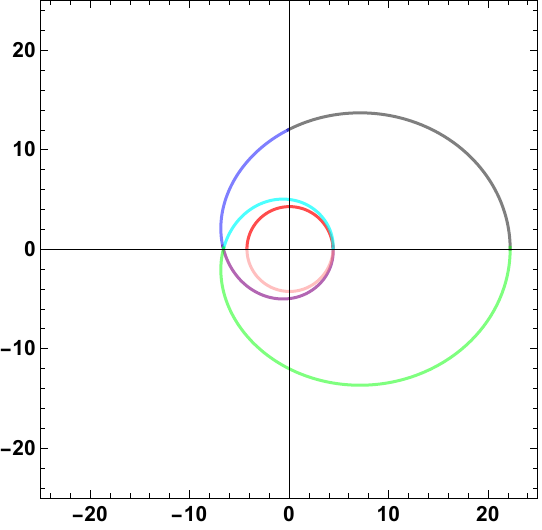} \\
			(a) Periodic orbit.
		\end{minipage} &
		\begin{minipage}{0.6\textwidth}
			\centering
			\includegraphics[width=0.8\linewidth]{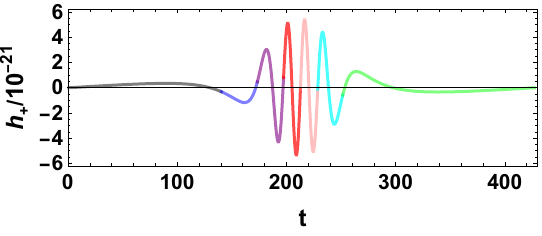} \\
			\vspace{10pt}
			\includegraphics[width=0.8\linewidth]{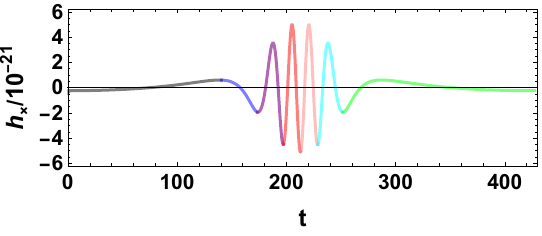} \\
			(b) Gravitational waveforms.
		\end{minipage}
	\end{tabular}
	\captionsetup{justification=raggedright,singlelinecheck=false}
	\caption{Gravitational waveform associated with the periodic orbit characterized by $q = (1;2;0)$. (a) Trajectory of the periodic orbit. (b) Corresponding gravitational waveform, with different segments highlighted in distinct colors. The parameters are set to $\alpha = 0.3$ and $\epsilon = 0.5$.}
	\label{GWcPO}
\end{figure*}

Figure \ref{GWcPO} employs distinct colours to show the correlation between the GW form and the periodic orbit, facilitating a thorough examination of the relationship between the periodic orbit and its corresponding gravitational radiation.
The waveform clearly encodes the zoom-whirl structure of the orbit over a full cycle, faithfully reflecting the alternating zooming and whirling segments of the small-mass object trajectory.
As shown in the figure, the signal amplitude reaches its maximum near perihelion and diminishes as the small object moves away from perihelion.
It follows that the smooth, slowly varying segments of the waveform correspond to the zoom phases, when the small object moves along a highly elongated elliptical path far from the BH, whereas the rapidly oscillating segments of the signal represent the whirl phases, arising when the object approaches the BH and executes tight circular motion near the horizon.
In the initial stages, when the object is distant from the BH and the gravitational field is comparatively weak, the wave is smooth and of low amplitude.
As the small object nears the BH, its trajectory gets increasingly twisted, the gravitational field intensifies, and both the amplitude and frequency of the GWs rise dramatically. 
In the concluding phases, as the particle approaches the event horizon, the GW signal reaches its highest point, marked by a sharp frequency sweep, a steep rise in amplitude, and increasingly pronounced oscillations.
As the small object accelerates and its orbit becomes progressively more distorted, the GW signal grows in both amplitude and frequency. These waves thus act as carriers of information about the object’s motion, the distribution of mass, and the surrounding spacetime, offering a powerful means to investigate the physical properties of BHs and their environments through waveform analysis.

\begin{figure*}
	\centering
	\includegraphics[width=7.6 cm]{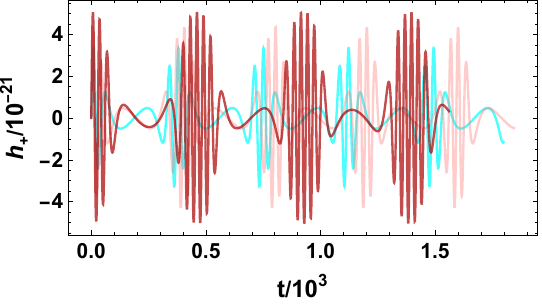}
	\includegraphics[width=7.6 cm]{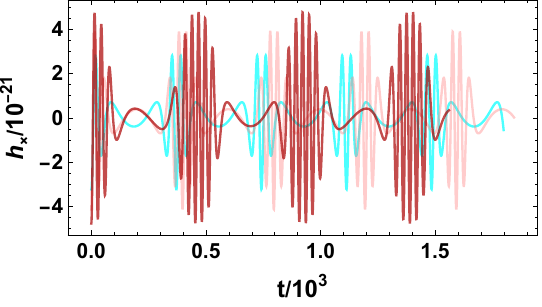}
	
	\parbox{\textwidth}{\centering (a) $\alpha=0.3$}
	\vspace{1pt} 
	
	\includegraphics[width=7.6 cm]{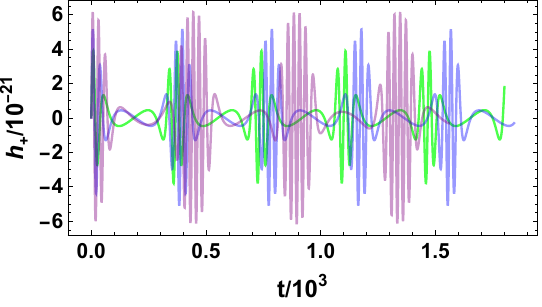}
	\includegraphics[width=7.6 cm]{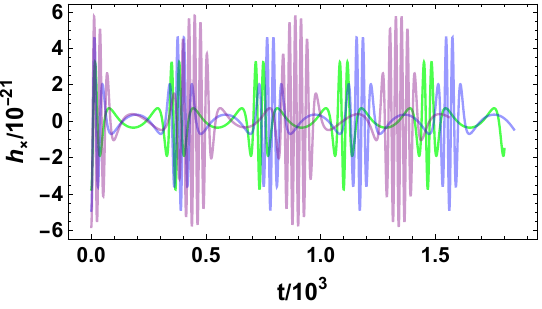}
	
	\parbox{\textwidth}{\centering (b) $\alpha=0.7$}
	
	\captionsetup{justification=raggedright,singlelinecheck=false}
	\caption{Gravitational waveforms arising from different periodic orbits for two values of $\alpha$ at fixed energy $E = 0.96$. Distinct orbits, labeled by $q = (z; w; \nu)$, are shown in different colors: $(1; 1; 0)$ in cyan and green, $(2; 1; 1)$ in pink and blue, and $(3; 2; 2)$ in red and purple. Panels (a) and (b) correspond to $\alpha = 0.3$ and $\alpha = 0.7$, respectively.}
	\label{wave1}
\end{figure*}

Figure \ref{wave1} shows the gravitational waveforms corresponding to increasing zoom numbers, with $(z, w, \nu) = (1;1;0), (2;1;1),$ and $(3;2;2)$. The results highlight a robust correlation between the waveform structure and the orbital motion of the small object. Each trajectory exhibits distinct zoom and whirl phases in the waveform that reflect the object’s trajectory. The number of smooth segments in the signal matches the number of orbital “leaves,” while the sharp oscillations correspond to the whirls in the corresponding orbit. Comparing the RSBH cases with $\alpha = 0.3$ and $\alpha = 0.7$, we find that the parameter $\alpha$ primarily influences the phase of the GW signal, while also producing a noticeable effect on its amplitude. Furthermore, orbits with larger zoom numbers $z$ generate waveforms with increasingly complex substructures, mirroring the greater number of “leaves” in the associated periodic orbit.

\begin{figure*}
	\centering
	\begin{tabular}{ c }
		\includegraphics[width=0.5\textwidth]{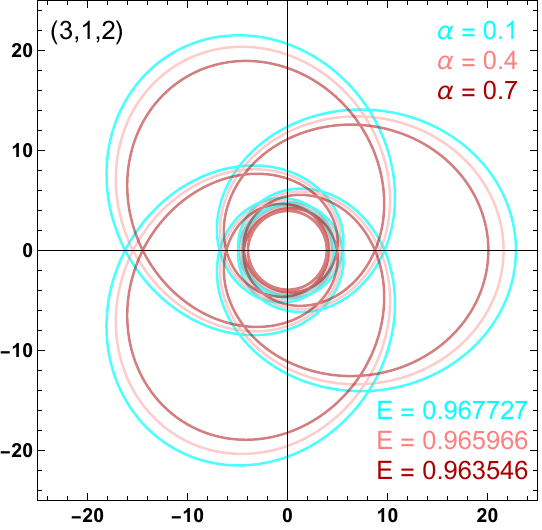}
	\end{tabular}%
	\begin{tabular}{cc}
		\includegraphics[width=0.45\textwidth]{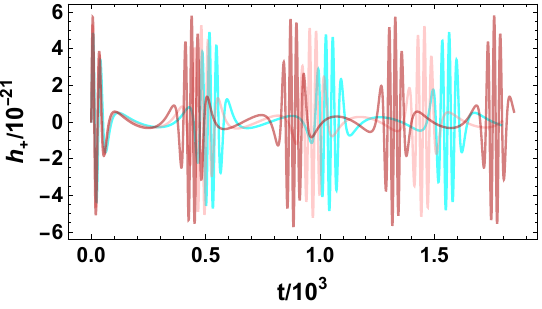}\\ 
		\includegraphics[width=0.45\textwidth]{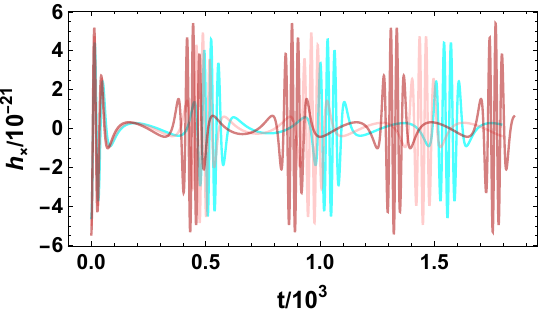} 
	\end{tabular}
	\captionsetup{justification=raggedright,singlelinecheck=false}
	\caption{The left panel shows a typical periodic orbit around an RSBH with $(z; w; \nu) = (3; 1; 2)$, obtained by fixing $L$ through $\epsilon = 0.5$, for three different values of $\alpha$ represented in distinct colors. The right panel displays the corresponding gravitational waveform for the EMRI system, consisting of a small object orbiting a supermassive RSBH}
	\label{wave2}
\end{figure*}

The RSBH model parameter $\alpha$ plays a significant role in shaping the gravitational waveforms generated by objects on periodic orbits. Considering the orbit $(z; w; \nu) = (3; 1; 2)$, as illustrated in Fig. \ref{wave2}, we find that increasing $\alpha$ leads to a shift in the phase of the GW signal along with a visible increase in amplitude.
In particular, a discernible phase difference is observed between the cases $\alpha = 0.1$ and $\alpha = 0.35$, which becomes even more apparent at $\alpha = 0.7$, along with visible amplitude variations.

\section{Discussion and conclusions}\label{SecC}

In this paper, we investigated the potential observational signatures of a class of regular BHs distinguished by a Minkowskian core and sub-Planckian curvature, concentrating on their periodic orbits and the GWs from EMRIs generated along such trajectories, in particular for $x = 2/3$ and $p = 2$.
We began with a brief review of the spherically symmetric regular BH solution constructed from an exponentially suppressed gravitational potential, followed by an analysis of its scalar invariants. The exponential factor in the metric ensures that the associated RSBH approaches a Minkowskian core in the limit $r \to 0$. The curvature of the geometry is asymptotically flat both at the core and at spatial infinity, attaining a finite maximum at an intermediate radius. 
We then examined the geodesic motion around the RSBH by analysing the effective potential and bound orbits for timelike particles, as well as the photon sphere and shadow silhouette for lightlike particles. We found that the model parameter $\alpha$ played a significant role in the near-horizon region, with a marked influence on the photon sphere radius. Thus, the shadow silhouette was affected, even when viewed from large distances.
Using the EHT observational data for M87* and Sgr A*, we determined that the parameter $\alpha$ was not subject to additional upper or lower observational bounds, as the predicted shadow radius consistently fell within the allowed intervals. In particular, the shadow radius exhibited only a slow dependence on $\alpha$, especially in the case of M87*. The shadow size did not exceed the $1\sigma$ or $2\sigma$ CLs. Taken together, these results indicated that $\alpha$ did not need to be restricted beyond the theoretical interval $\alpha \in [0, 2 M^{4/3}/e]$, required to ensure a singularity-free BH geometry.
Future observations of other central BHs at curvature scales different from those of M87* and Sgr A* will provide independent tests, offering improved opportunities to constrain modified gravity models and refine predictions of spacetime properties from captured images. Such data would open the way to modeling central BHs as RSBHs with (p = 2), thereby reinforcing their viability as astrophysical BH candidates within the precision of forthcoming observational data.

We also arrived at the equations of motion from the Lagrangian and determined the associated effective potential for massive test particles. Our analysis showed that the barrier height of the effective potential diminished as the dimensionless deviation parameter $\alpha$ increased. By examining this potential, we numerically traced the dependence of $\alpha$ on the properties of the MBO and ISCO. Specifically, both the radius and orbital angular momentum of the MBO decreased with increasing $\alpha$, while for the ISCO, the radius, orbital angular momentum $L$, and energy $E$ all exhibited the same downward trend. In addition, we mapped the allowed parameter region in $(L , E)$ of angular momentum and energy for bound orbits around the RSBH and found that bound orbits with fixed angular momentum and larger values of $\alpha$ admitted a higher upper energy bound.

From Tables \ref{Tab:EList} and \ref{Tab:LList}, we found that for periodic orbits characterized by the same rational number $q$, both the orbital energy and angular momentum consistently decreased as the deviation parameter $\alpha$ increased. These findings may offer a means to distinguish an RSBH from both the Schwarzschild and Bardeen black holes, which share the same asymptotic behavior at infinity, through the analysis of periodic orbits around the central source.

Next, we explored an EMRI system composed of a stellar-mass object with $m = 10M_{\odot}$ orbiting a supermassive RSBH of mass $M = 10^{6}M_{\odot}$. Employing the numerical kludge approach, we studied the corresponding gravitational waveforms by placing the EMRI system at a luminosity distance of $D_{L} = 200 \, \mathrm{Mpc}$ from the detector, with the inclination angle $\iota = \pi/4$ and longitude of pericenter $\zeta = \pi/4$. Our results revealed a pronounced connection between the waveforms emitted by the orbiting object and its zoom–whirl dynamics.
Specifically, the polarizations $h_{+}$ and $h_{\times}$ exhibited quiet phases during the elongated zoom phases, followed by pronounced bursts during the roughly circular whirl phases. Moreover, orbits with higher zoom numbers were associated with increasingly intricate waveform substructures.
These characteristics indicate that future high-precision GW signatures spanning the entire zoom-whirl cycle could be used to identify the zoom-whirl structure of periodic orbits and, in turn, reconstruct the particle’s trajectory. In addition, the corrections owing to quantum gravity effects had a pronounced influence on the waveforms: increasing the model parameter $\alpha$, which arises from quantum effects of gravity and encodes modifications to the standard Heisenberg uncertainty principle owing to quantum gravity, led to both a phase shift and an amplitude rise in the GWs.
Furthermore, the phase of the GW signals associated with periodic orbits served as a sensitive probe for constraining the parameters of the underlying BH solutions. These features could be employed to identify the orbital structure of EMRIs and to assess the future GW detectors in testing RSBH models.

\acknowledgments
The author would like to thank Sumarna Haroon from Zhejiang University of Technology.
The research of L.M.N., S.Z. and H.H. was supported by 
the European Union.-Next Generation UE/MICIU/Plan de Recuperacion, Transformacion y Resiliencia/Junta de Castilla y Leon, 
RED2022-134301-T financed by MICIU/AEI/10.13039/501100011033, 
and PID2020-113406GB-I00 financed by MICIU/AEI/10.13039/501100011033. 
T.Z. and S.L are supported by the National Natural Science Foundation of China under Grants No.~12275238, the Zhejiang Provincial Natural Science Foundation of China under Grants No.~LR21A050001 and No.~LY20A050002, the National Key Research and Development Program of China under Grant No. 2020YFC2201503, and the Fundamental Research Funds for the Provincial Universities of Zhejiang in China under Grant No.~RF-A2019015.


\begin{thebibliography}{99}

\bibitem{AbbottPRL2016-1}
B.P. Abbott et al. [LIGO Scientific and Virgo Collaborations], Phys. Rev. Lett. \textbf{116} (2016) 061102.

\bibitem{AbbottPRL2016-2}
B. P. Abbott et al. [LIGO Scientific and Virgo Collaborations], Phys. Rev. Lett. \textbf{116} (2016) 241102.

\bibitem{AbbottPRX2021}
R. Abbott et al. (LIGO Scientific and Virgo Collaborations), Phys. Rev. X \textbf{11} (2021) 021053.


\bibitem{AkiyamaL12019}
K. Akiyama, et al., (Event Horizon Telescope), Astrophys. J. Lett. \textbf{875} (2019) L1.
\bibitem{AkiyamaL52019}
K. Akiyama, et al., (Event Horizon Telescope), Astrophys. J. Lett. \textbf{875} (2019) L5.
\bibitem{AkiyamaL62019}
K. Akiyama, et al., (Event Horizon Telescope), Astrophys. J. Lett. \textbf{875} (2019) L6.


\bibitem{AkiyamaL122022} 
K. Akiyama, et al., (Event Horizon Telescope Collaboration), Astrophys. J. Lett. \textbf{930} (2022) L12. 
\bibitem{AkiyamaL172022} 
K. Akiyama, et al., (Event Horizon Telescope Collaboration), Astrophys. J. Lett. \textbf{930} (2022) L17. 

\bibitem{PsaltisPRL2020}
D. Psaltis et al. (Event Horizon Telescope Collaboration), Phys. Rev. Lett. \textbf{125} (2020) 141104.

\bibitem{GRAVITYAA2020}
R. Abuter et al. (GRAVITY Collaboration), Astron. Astrophys. \textbf{636} (2020) L5.

\bibitem{BaubockAA2020}
M. Baub\"{o}ck et al. (GRAVITY Collaboration), Astron. Astrophys. \textbf{635} (2020) A143.

\bibitem{BarackCQG2019} 
L. Barack et al.,  Class. Quant. Grav. \textbf{36} (2019) 143001.

\bibitem{OppenheimerPR1930}
J. R. Oppenheimer and H. Snyder, Phys. Rev. \textbf{56} (1939) 455.


\bibitem{LuminetAA1979}
J. P. Luminet, Astron. Astrophys. \textbf{75} (1979) 228.

\bibitem{FalckeApJL13}
H. Falcke, and F. Melia, E. Agol. Astrophys. J. Lett. 
\textbf{528} (2000) L13.  		

\bibitem{GrallaPRD2019}
S. E. Gralla, D. E. Holz, and R. M. Wald, Phys. Rev. D \textbf{100} (2019) 024018.

\bibitem{CunhaGRG2018} 
P. V. P. Cunha, and C. A. R. Herdeiro, Gen. Relativ. Gravitation \textbf{50} (2018) 42. 	

\bibitem{CunhaPLB2017}
P. V. P. Cunha, C. A. R. Herdeiro, B. Kleihaus, J. Kunz, and E. Radu, Phys. Lett. B \textbf{768} (2017) 373.

\bibitem{NarayanApJL2019} 
R. Narayan, M. D. Johnson, and C. F. Gammie, Astrophys. J. Lett. \textbf{885} (2019) L33.

\bibitem{PerlickPR2022}
V. Perlick, and O. Yu. Tsupko, Phys. Rep. \textbf{947} (2022) 1.	


\bibitem{TsupkoPRD2017}
O. Y. Tsupko, Phys. Rev. D \textbf{95} (2017) 104058.
\bibitem{BroderickApJ2022}
A. E. Broderick, D. W. Pesce, P. Tiede, H. Y. Pu, R. Gold, R. Anantua, S. Britzen, C. Ceccobello, K. Chatterjee, Y. Chen et al., Astrophys. J. \textbf{935} (2022) 61. 
\bibitem{MizunoNatAst2018}
Y. Mizuno, Z. Younsi, C. M. Fromm, O. Porth, M. De Laurentis, H. Olivares, H. Falcke, M. Kramer, L. Rezzolla, Nat. Astron. \textbf{2} (2018) 585.



\bibitem{AtamurotovPRD2013}
F. Atamurotov, A. Abdujabbarov, and B. Ahmedov,  Phys. Rev. D \textbf{88} (2013) 064004.
\bibitem{AbdujabbarovSS2016}
A. Abdujabbarov, B. Juraev, B. Ahmedov, and Z. Stuchlik,  Astrophys. Space Sci. \textbf{361} (2016) 226. 
\bibitem{AbdikamalovPRD2019}
A. B. Abdikamalov, A. A. Abdujabbarov, D. Ayzenberg, D. Mala-farina, C. Bambi, and B. Ahmedov, Phys. Rev. D \textbf{100} (2019) 024014. 
\bibitem{AtamurotovPRD2015}
F. Atamurotov, and B. Ahmedov,  Phys. Rev. D \textbf{92} (2015) 084005.
\bibitem{AtamurotovCPC2023}
F. Atamurotov, I. Hussain, G. Mustafa, and A. \"{O}vg\"{u}n, Chin. Phys. C \textbf{47} (2023) 025102. 
\bibitem{BelhajPLB2021}
A. Belhaj, H. Belmahi, M. Benali, W. El Hadri, H. El Moumni, and E. Torrente-Lujan,  Phys. Lett. B \textbf{812} (2021) 136025. 
\bibitem{BelhajCQG2021}
A. Belhaj, M. Benali, A. El Balali, H. El Moumni, and S. E. Ennadifi, Class. Quant. Grav. \textbf{37} (2020) 215004. 

\bibitem{WeiJCAP2019}
S.-W. Wei, Y.-C. Zou, Y.-X. Liu, and R.B. Mann,  J. Cosmol. Astropart. Phys. \textbf{08} (2019) 030.
\bibitem{LingPRD2021}
R. Ling, H. Guo, H. Liu, X.-M. Kuang, and B. Wang,  Phys. Rev. D \textbf{104} (2021) 104003. 
\bibitem{TsukamotoPRD2018}
N. Tsukamoto, Phys. Rev. D  \textbf{97} (2018) 064021 . 

\bibitem{AraujoFilhoCQG2024}
A. A. Ara\'{u}jo Filho, H. Hassanabadi, N. Heidari, J. K\v{r}\'{i}\v{z}, and S. Zare,  Class. Quant. Grav. \textbf{41} (2024) 055003.

\bibitem{RayimbaevPoDU2022}
J. Rayimbaev, B. Majeed, M. Jamil, K. Jusufi, and A. Wang, Phys. Dark Univ. \textbf{35} (2022) 100930.

\bibitem{PerlickPRD2018}
V. Perlick, O. Y. Tsupko, and G. S. Bisnovatyi-Kogan, Phys. Rev. D \textbf{97} (2018) 104062. 

\bibitem{RosaPRD2023}
J. L. Rosa, Phys. Rev. D \textbf{107} (2023) 084048.


\bibitem{VagnozziCQG2023}
S. Vagnozzi et al., Class. Quant. Grav. \textbf{40} (2023) 165007. 
\bibitem{KocherlakotaPRD2021} 
P. Kocherlakota et al., Phys. Rev. D \textbf{103} (2021) 104047. 



\bibitem{UniyalPoDU2023}
A. Uniyal,  R. C. Pantig, and A. \"{O}vg\"{u}n, Phys. Dark Univ. \textbf{40} (2023) 101178.  	


\bibitem{LambiaseEPJC2023}
G. Lambiase, R. C. Pantig, D. J. Gogoi, and Ali Övgün, Eur. Phys. J. C \textbf{83} (2023) 679.
\bibitem{KhodadiPRD2022}
M. Khodadi, and G. Lambiase, Phys. Rev. D \textbf{106} (2022) 104050.
\bibitem{KhodadiJCAP2021}
M. Khodadi, G. Lambiase, and D.F. Mota, J. Cosmol. Astropart. Phys. \textbf{09} (2021) 028.
\bibitem{PanotopoulosPRD2021}
G. Panotopoulos, \'{A}. Rinc\'{o}n, and I. Lopes,  Phys. Rev. D  \textbf{103} (2021) 104040.
\bibitem{EslamPanahEPJC2024}
B. Eslam Panah, S. Zare, and H. Hassanabadi, Eur. Phys. J. C \textbf{84} (2024) 259.


\bibitem{Meng2025arxiv}
Y. Meng, X-J. Wang, Y.-Z. Li, and X.-M. Kuang, arXiv:2501.02496v1

\bibitem{Xu2024arxiv}
Y. Xu, H. Huang, M.-Y. Lai, and D.-C. Zou, arXiv:2407.06562v2

\bibitem{Huang2025arxiv}
H. Huang, Y. Xu, M.-Y. Lai, and D.-C. Zou, "Distinguishing Black Holes from String-Inspired Euler-Heisenberg Theory Through Shadow Images." (2025).

\bibitem{FengAstpP2025}
H. Feng, R. J. Yang, and W. Q. Chen, Astropart. Phys. \textbf{166} (2025) 103075.

\bibitem{Waseem2025arxiv}
H. Waseem, N. J. L. S. Lobos, A. \"{O}vg\"{u}n, and R. C. Pantig, Eur. Phys. J. C \textbf{85} (2025) 629. 

\bibitem{Kala2025arxiv}
S. Kala, H. Nandan, K. Maithani, S. Roy, and A. Abebe, arXiv:2503.19571

\bibitem{Yin2025arxiv}
J. Yin, T. Y. He, M. Liu, B. Chen, Z. W. Han, and R. J. Yang, Nucl. Phys. B. \textbf{1018} (2025) 117004. 

\bibitem{ZareJCAP2024}
S. Zare, L. M. Nieto, X.-H. Feng, S.-H. Dong and H. Hassanabadi, JCAP \textbf{08} (2024) 041.

\bibitem{StuchlikEPJC2019}
Z. Stuchl\'{i}k, and J. Schee, Eur. Phys. J. C \textbf{79} (2019) 44. 


\bibitem{WuPoDU2024}
S. R. Wu, B. Q. Wang, Z. W. Long, and Hao Chen, Phys. Dark Univ. \textbf{44} (2024) 101455.

\bibitem{PantigJCAP2022} 	
R.C. Pantig, and A. \"{O}vg\"{u}n, J. Cosmol. Astropart. Phys. \textbf{08} (2022) 056 . 

\bibitem{ChenEPJP2024}
H. Chen, S.-H. Dong, E. Maghsoodi, S. Hassanabadi, J. K\v{r}\'{i}\v{z}, S. Zare, and H. Hassanabadi, Eur. Phys. J. Plus \textbf{139} (2024) 759.
	
\bibitem{CapozzielloJCAP2023}
S. Capozziello, S. Zare, D.F. Mota, and H. Hassanabadi, J. Cosmol. Astropart. Phys.  \textbf{05} (2023) 027.
\bibitem{CapozzielloPoDU2023}
S. Capozziello, S. Zare, L. M. Nieto, and H. Hassanabadi,Phys. Dark Univ. \textbf{50} (2025) 102065.
\bibitem{XuJCAP2018}
Z. Xu, X. Hou, X. Gong, and J. Wang, J. Cosmol. Astropart. Phys. \textbf{09} (2018) 038.	

\bibitem{Nieto2025}
L. M. Nieto, F. Hosseinifar, K. Boshkayev, S. Zare, and H. Hassanabadi, 	arXiv:2507.14305.
\bibitem{SekhmaniJHEA2025}
Y. Sekhmani, S. Zare, L. M. Nieto, H. Hassanabadi, and K Boshkayev, JHEAp \textbf{47} (2025) 100389.

\bibitem{KonoplyaPLB2019}
R. A. Konoplya, Phys. Lett. B \textbf{795} (2019) 1. 

\bibitem{ZarePLB2024}
S. Zare, L. M. Nieto, F. Hosseinifar, X.-H. Feng, and H. Hassanabadi, Phys. Lett. B \textbf{859} (2024) 139125.


\bibitem{PenrosePRL1965} R. Penrose, Phys. Rev. Lett. \textbf{14} (1965) 57. 
\bibitem{Penrose1969} R. Penrose, Riv. Nuovo Cimento \textbf{1} (1969) 252.
\bibitem{Hawking1966} S. Hawking, Proc. R. Soc. A \textbf{294} (1966) 511.

\bibitem{Joshi2011} P. S. Joshi and D. Malafarina, Int. J. Mod. Phys. D \textbf{20} (2011) 2641.
\bibitem{GoswamiPRL2006} R. Goswami, P. S. Joshi, and P. Singh, Phys. Rev. Lett. \textbf{96} (2006) 031302.
\bibitem{JanisPRL1968} A. I. Janis, E. T. Newman, and J. Winicour, Phys. Rev. Lett. \textbf{20} (1968) 878.


\bibitem{HawkingPRD1976} S. W. Hawking, Phys. Rev. D \textbf{14} (1976) 2460.
\bibitem{Giddings1992} S. B. Giddings, Phys. Rev. D \textbf{46} (1992) 1347.
\bibitem{Hawking1975} S. W. Hawking, Commun. Math. Phys. \textbf{43} (1975) 199.
\bibitem{Preskill} J. Preskill, arXiv:hep-th/9209058.
\bibitem{LiIJMPD2013} X. Li, Y. Ling, and Y. G. Shen, Int. J. Mod. Phys. D \textbf{22} (2013) 1342016.
\bibitem{ChenPR2015} P. Chen, Y. C. Ong, and D. h. Yeom, Phys. Rep. \textbf{603} (2015) 1. 
\bibitem{CasadioPLB2000} R. Casadio and B. Harms, Phys. Lett. B \textbf{487} (2000) 209.


\bibitem{DeWittPR1967} B. S. DeWitt, Phys. Rev. \textbf{160} (1967) 1113.
\bibitem{tHooftNPB1985} G.\,{}'t Hooft, Nucl. Phys. B \textbf{256} (1985) 727.
\bibitem{CallanPRD1992} C. G. Callan, Jr., S. B. Giddings, J. A. Harvey, and A. Strominger, Phys. Rev. D \textbf{45} (1992) R1005.
\bibitem{DonoghuePRL1994} J. F. Donoghue, Phys. Rev. Lett. \textbf{72} (1994) 2996.
\bibitem{GarayIJMPA1995} L. J. Garay, Int. J. Mod. Phys. A \textbf{10} (1995) 145.
\bibitem{HanPLB2005} T. Han, and S. Willenbrock, Phys. Lett. B \textbf{616} (2005) 215. 
\bibitem{CalmetEPJC77} X. Calmet, and B. K. El-Menoufi, Eur. Phys. J. C \textbf{77} (2017) 243.

\bibitem{Ashtekar2023} A. Ashtekar, J. Olmedo, and P. Singh, arXiv:2301.01309. 
\bibitem{LanIJTP2023} C. Lan, H. Yang, Y. Guo, and Y. G. Miao, Int. J. Theor. Phys. \textbf{62} (2023) 202. 

\bibitem{Bardeen1968} J. M. Bardeen, in Proceeding of the International Conference GR5 (1968), Tbilisi, USSR, Georgia, pp. 174–180.

\bibitem{HaywardPRL2006}
S. A. Hayward, Phys. Rev. Lett. \textbf{96} (2006) 031103.
\bibitem{FrolovJHEP2014} 
V. P. Frolov, J. High Energy Phys. \textbf{05} (2014) 049.


	
\bibitem{LingCQG2023}
Y. Ling, M.-H. Wu, Class. Quant. Grav. \textbf{40} (2023) 075009.



\bibitem{Culetu} H. Culetu, arXiv:1305.5964.
\bibitem{CuletuIJTP2015} H. Culetu, Int. J. Theor. Phys. \textbf{54} (2015) 2855.
\bibitem{RodriguesPRD2016} M. E. Rodrigues, E. L. B. Junior, G. T. Marques, and V. T. Zanchin, Phys. Rev. D \textbf{94} (2016) 024062.
\bibitem{SimpsonUniv2019} A. Simpson and M. Visser, Universe \textbf{6} (2019) 8.
\bibitem{ZengPRD2023}
W. Zeng, Y. Ling, Q.-Q. Jiang, and G.-P. Li, Phys. Rev. D \textbf{108} (2023) 104072.
\bibitem{TangEPJC2024}
C. Tang,Y. Ling, Q.-Q. Jiang, and G.-P. Li, Eur. Phys. J. C, \textbf{84} (2024) 1296.
\bibitem{ZhangEPJC2024}
D. Zhang, H. Gong, G. Fu, J.-P. Wu and Q. Pan, Eur. Phys. J. C \textbf{84} (2024) 564.	



\bibitem{LISA1}
P. Amaro-Seoane et al. (LISA Collaboration), Laser Inter-ferometer Space Antenna, arXiv:1702.00786.
\bibitem{LISA2}
P. A. Seoane et al. (LISA Collaboration), Astrophysics with the Laser Interferometer Space Antenna, Living Rev. Relativity \textbf{26}, (2023) 2.

\bibitem{HuNSR2017}
W.-R. Hu and Y.-L. Wu, Natl. Sci. Rev. \textbf{4} (2017) 685.

\bibitem{LuoCQG2016}
J. Luo et al. (TianQin Collaboration), Class. Quant. Grav. \textbf{33} (2016) 035010.

\bibitem{DECIGO}
M. Musha (DECIGO Working Group), Proc. SPIE Int. Soc. Opt. Eng. \textbf{10562} (2017) 105623T.


\bibitem{HughesCQG2001}
S. A. Hughes, Class. Quant. Grav. \textbf{18} (2001) 4067.

\bibitem{GlampedakisCQG2005-2}
K. Glampedakis, Class. Quant. Grav. \textbf{22} (2005) S605.

\bibitem{BarausseGRG2020} 
E. Barausse, E. Berti, T. Hertog, S. A. Hughes, P. Jetzer, P. Pani, T. P. Sotiriou, N. Tamanini, H. Witek, K. Yagi et al.,  Gen. Relativ. Gravit. \textbf{52} (2020) 81.


\bibitem{CardosoLRR2019} 
V. Cardoso and P. Pani, Living Rev. Relativity \textbf{22} (2019) 4. 

\bibitem{BabakPRD2017} 
S. Babak, J. Gair, A. Sesana, E. Barausse, C. F. Sopuerta, C. P. L. Berry, E. Berti, P. Amaro-Seoane, A. Petiteau, and A. Klein, Phys. Rev. D \textbf{95} (2017) 103012.

















\bibitem{LevinPRD2009}
 J. Levin and B. Grossman, Phys. Rev. D \textbf{79} (2009) 043016.
 
 \bibitem{LevinPRD2008}
 J. Levin, and G. Perez-Giz, Phys. Rev. D \textbf{77} (2008) 103005.
 
 \bibitem{LevinCQG2009}
 J. Levin, Class. Quant. Grav. \textbf{26} (2009) 235010.

\bibitem{GrossmanPRD2009}
R. Grossman, and J. Levin, Phys. Rev. D \textbf{79} (2009) 043017.

\bibitem{MisraPRD2010}
V. Misra, and J. Levin, Phys. Rev. D \textbf{82} (2010) 083001.

\bibitem{BabarPRD2017}
G. Z. Babar, A. Z. Babar, and Y.-K. Lim, Phys. Rev. D \textbf{96} (2017) 084052.


\bibitem{HaroonPRD2025}
S. Haroon, T. Zhu,  Phys. Rev. D \textbf{112} (2025) 044046.

\bibitem{Alloqulov2025arxiv}
M. Alloqulov, T. Xamidov, S. Shaymatov, and B. Ahmedov, Eur. Phys. J. C  \textbf{85} (2025) 798.



\bibitem{GaoAP2020}
B. Gao, and X.-M. Deng, Ann. Phys. \textbf{418} (2020) 168194.

\bibitem{MengEPJC2025}
L. Meng, Z. Xu, and M. Tang, Eur. Phys. J. C \textbf{85} (2025) 306.

\bibitem{WangJCAP2025}
C.-H. Wang, X.-C. Meng, Y.-P. Zhang, T. Zhu, S.-W. Wei, JCAP \textbf{07} (2025) 021.

\bibitem{LiuCTP2019}
C. Liu, C. Ding and J. Jing, Commun. Theor. Phys. \textbf{71} (2019) 1461.

\bibitem{TuPRD2023}
Z.-Y. Tu, T. Zhu, and A. Wang, Phys. Rev. D \textbf{108} (2023) 024035.

\bibitem{LinPoDU2021}
H.-Y. Lin, and X.-M. Deng, Phys. Dark Univ. \textbf{31} (2021) 100745. 

\bibitem{DengPoDU2020}
X.-M. Deng, Phys. Dark Univ.  \textbf{30} (2020) 100629.

\bibitem{YangJCAP2025}
S. Yang, Y.-P. Zhang, T. Zhu, L. Zhao, and Y.-X. Liu, JCAP \textbf{01} (2025) 091.

\bibitem{WeiPRD2019}
S.-W. Wei, J. Yang, and Y.-X. Liu, Phys. Rev. D \textbf{99} (2019) 104016.

\bibitem{DengEPJC2020}
X.-M. Deng, Eur. Phys. J. C \textbf{80} (2020) 489.

\bibitem{LinEPJC2023}
H.-Y. Lin, and X.-M. Deng, Eur. Phys. J. C \textbf{83} (2023) 311.









\bibitem{LinAP2023} 
H. Y. Lin, and X. M. Deng, Ann. Phys. (N.Y.) \textbf{455} (2023) 169360.

\bibitem{ChanGRG2025} 
Z. C. S. Chan, and Y. K. Lim, Gen. Relativ. Gravit. \textbf{57} (2025) 35.

\bibitem{ZhouEPJC2020}
T.-Y. Zhou, and Y. Xie, Eur. Phys. J. C \textbf{80} (2020) 1070.

\bibitem{ZhangEPJC2022}
J. Zhang, and Y. Xie, Eur. Phys. J. C \textbf{82} (2022) 854.

\bibitem{YaoPRD2023}
J.-T. Yao, and X. Li, Phys. Rev. D \textbf{108} (2023) 084067.

\bibitem{QiEPJC2024}
Q. Qi, X.-M. Kuang, Y.-Z. Li and Y. Sang,  Eur. Phys. J. C
\textbf{84} (2024) 645.

\bibitem{Li-KuangPRD2023}
Y.-Z. Li, and X.-M. Kuang, Phys. Rev. D \textbf{107} (2023) 064052.

\bibitem{JiangPoDU2024}
H. Jiang, M. Alloqulov, Q. Wu, S. Shaymatov, and T. Zhu, Phys. Dark Univ. \textbf{46} (2024) 101627.

\bibitem{Lu2025arxiv}
S. Lu, and T. Zhu, arXiv:2505.00294v1.

\bibitem{HuangPRD2025}
L. Huang, Phys. Rev. D \textbf{111} (2025) 084038.

\bibitem{ZhaoEPJC2024}
L. Zhao, M. Tang, and Z. Xu, Eur. Phys. J. C \textbf{85} (2025) 36.

\bibitem{ShabbirPoDU2025}
O. Shabbir, M. Jamil, and M. Azreg-A\"{\i}nou, Phys. Dark Univ. \textbf{47} (2025) 101816.

\bibitem{LiEPJC2024}
Y.-Z. Li, X.-M. Kuang, and Y. Sang,  Eur. Phys. J. C \textbf{84} (2024) 529. 

\bibitem{Wang:2025wob}
C.~H.~Wang, Y.~P.~Zhang, T.~Zhu and S.~W.~Wei,
[arXiv:2508.20558 [gr-qc]].



\bibitem{Chandrasekhar} S. Chandrasekhar, ``The Mathematical Theory of Black Holes" (Oxford University Press, New York, 1998).

\bibitem{Misner}
C. W. Misner, K.S. Thorne and J.A. Wheeler, Gravitation, W.H. Freeman, U.S.A. (1973) [ISBN:978-0-7167-0344-0, 978-0-691-17779-3].


\bibitem{BabakPRD2007}
S. Babak, H. Fang, J.R. Gair, K. Glampedakis, and S.A. Hughes, Phys.Rev.D \textbf{75} (2007) 024005











\bibitem{XiangIJMPD2013}
L. Xiang, Y. Ling, and Y.G. Shen,  Int. J. Mod. Phys. D \textbf{22}  (2013) 1342016.

\bibitem{LiAP2018}
X. Li, Y. Ling, Y.-G. Shen, C.-Z. Liu, H.-S. He, and L.-F. Xu,  Ann. Phys. \textbf{396} (2018) 334.


\bibitem{Straumann}
N. Straumann, General Relativity with Applications to Astrophysics (Springer, New York, 2004).






\bibitem{HuEPJC2022}
S. Hu, C. Deng, D. Li, X. Wu, and E. Liang,  Eur. Phys. J. C \textbf{82} (2022) 885.

\bibitem{LiEPJC2021}
G.-P. Li, and K.-J. He, Eur. Phys. J. C \textbf{81} (2021) 1018.














\bibitem{HughesPRD2000}
S.~A.~Hughes, Phys. Rev. D \textbf{61}  (2000) 084004.

\bibitem{HughesPRD2001}
S.~A.~Hughes, Phys. Rev. D \textbf{64} (2001) 064004.

\bibitem{GlampedakisPRD2002}
K.~Glampedakis, and D.~Kennefick, Phys. Rev. D \textbf{66} (2002) 044002.
	
\bibitem{HughesPRL2005}
S.~A.~Hughes, S.~Drasco, E.~E.~Flanagan, and J.~Franklin, Phys. Rev. Lett. \textbf{94} (2005) 221101.
	
\bibitem{DrascoCQG2005}
S.~Drasco, E.~E.~Flanagan, and S.~A.~Hughes, Class. Quant. Grav. \textbf{22} (2005) S801.

\bibitem{GairPRD2005}
J.~R.~Gair, and K.~Glampedakis, Phys. Rev. D \textbf{73} (2006) 064037.

\bibitem{GlampedakisCQG2005}
K.~Glampedakis, and S.~Babak, Class. Quant. Grav. \textbf{23} (2006) 4167.


\bibitem{DrascoPRD2005}
S.~Drasco and S.~A.~Hughes, Phys. Rev. D \textbf{73} (2006) 024027.

\bibitem{SundararajanPRD2007}
P.~A.~Sundararajan, G.~Khanna, and S.~A.~Hughes, Phys. Rev. D \textbf{76} (2007) 104005.

\bibitem{SundararajanPRD2008}
P.~A.~Sundararajan, G.~Khanna, S.~A.~Hughes, and S.~Drasco, Phys. Rev. D \textbf{78} (2008) 024022.

\bibitem{MillerPRD2021}
J.~Miller and A.~Pound, Phys. Rev. D \textbf{103} (2021) 064048.

\bibitem{IsoyamaPRL2022}
S.~Isoyama, R.~Fujita, A.~J.~K.~Chua, H.~Nakano, A.~Pound, and N.~Sago, Phys. Rev. Lett. \textbf{128} (2022) 231101.


\bibitem{ZiPLB2024}
T. Zi, Phys. Lett. B \textbf{850} (2024) 138538

\bibitem{ZiPRD2024}
T. Zi, and Peng-Cheng Li, Phys. Rev. D \textbf{109} (2024) 064089.


(Erratum: Phys. Rev. D \textbf{77} (2008) 049902). 

\bibitem{ThornePMP1980}
 K. S. Thorne, Rev. Mod. Phys. \textbf{52} (1980) 299.
 



















\end{thebibliography}
\end{document}